\documentclass[accepted]{arXiv2025} % for initial submission
\usepackage{natbib} % has a nice set of citation styles and commands
\usepackage{mathtools} % amsmath with fixes and additions
\usepackage{booktabs} % commands to create good-looking tables
\usepackage{tikz} % nice language for creating drawings and diagrams

\usepackage{amsmath,amsfonts}

\def\eqref#1{equation~\ref{#1}}
\def\1{\bm{1}}

\DeclareMathAlphabet{\mathsfit}{\encodingdefault}{\sfdefault}{m}{sl}
\SetMathAlphabet{\mathsfit}{bold}{\encodingdefault}{\sfdefault}{bx}{n}

\newcommand{\bfI}{{\bf I}}

\newcommand{\bfS}{{\bf S}}

\newcommand{\bfy}{{\bf y}}
\newcommand{\bfu}{{\bf u}}
\newcommand{\bfq}{{\bf q}}
\newcommand{\bfp}{{\bf p}}

\newcommand{\bfv}{{\bf v}}

\newcommand{\bfz}{{\bf z}}

\newcommand{\hf}{{\frac 12}}

\newcommand{\bfepsilon}{{\boldsymbol \epsilon}}
\newcommand{\bftheta}{{\boldsymbol \theta}}

\usepackage[english]{babel}
\usepackage[controls]{animate}
\usepackage{tikz}
\usetikzlibrary{arrows}
\usepackage{algorithm}
\usepackage{algpseudocode}
\usepackage{amsmath}
\usepackage{amsfonts}
\usepackage{graphicx}

\newtheorem{definition}{Definition}
\newtheorem{example}{Example}

\newtheorem{comment}{Comment}

\usepackage[utf8]{inputenc} % allow utf-8 input
\usepackage[T1]{fontenc}    % use 8-bit T1 fonts
\usepackage{hyperref}       % hyperlinks
\usepackage{url}            % simple URL typesetting
\usepackage{booktabs}       % professional-quality tables
\usepackage{amsfonts}       % blackboard math symbols
\usepackage{nicefrac}       % compact symbols for 1/2, etc.
\usepackage{microtype}      % microtypography
\usepackage{xcolor}         % colors
\usepackage{cleveref}

\title{Probabilistic Forecasting for Dynamical Systems with Missing or Imperfect Data }

\author[1,2]{Siddharth Rout}
\author[1,2]{Eldad Haber}
\author[3]{St\'ephane Gaudreault}
\affil[1]{%
    Institute of Applied Mathematics\\
    University of British Columbia\\
    Vancouver, BC, Canada
}
\affil[2]{%
    Department of Earth, Ocean and Atmospheric Sciences\\
    University of British Columbia\\
    Vancouver, BC, Canada
}
\affil[3]{%
    Recherche en pr\'evision num\'erique atmosph\'erique\\ Environnement et Changement climatique Canada \\
    Dorval, QC, Canada
}
  
\begin{document}
\maketitle

\begin{abstract}
  The modeling of dynamical systems is essential in many fields, but applying machine learning techniques is often challenging due to incomplete or noisy data. This study introduces a variant of stochastic interpolation (SI) for probabilistic forecasting, estimating future states as distributions rather than single-point predictions. We explore its mathematical foundations and demonstrate its effectiveness on various dynamical systems, including the challenging WeatherBench dataset.
\end{abstract}

\section{Introduction}\label{sec:intro}

The modeling of dynamical systems is of paramount importance across numerous scientific and engineering disciplines, including weather prediction, climate science, finance, and biology \citep{brin2002introduction, tu2012dynamical}. Traditional methods for predicting the behavior of these systems often rely on explicit mathematical models derived from known physical laws.
Such approaches are well summarized in \cite{reich2015probabilistic} and typically employ a combination of numerical methods, Bayesian inference, and sensitivity analysis.
However, these methods can be limited by their underlying assumptions and the complexity of the systems they aim to model \citep{guckenheimer1993dynamical, guckenheimer2013nonlinear}. In particular, they require a high-fidelity model with sufficiently high resolution and accurate input data. Furthermore, such approaches often demand significant computational resources, making them difficult to compute within a reasonable time frame \citep{moin1998direct, lynch2008origins, wedi2015modelling}.

In recent years, machine learning (ML) has emerged as a powerful tool for predicting dynamical systems, leveraging its ability to learn patterns and relationships from data without requiring explicit knowledge of the underlying physical laws \citep{Ghadami2022, pathak2022fourcastnet, lam2022graphcast, bodnar2024aurora}. ML approaches are well suited for this task, as they can automatically identify patterns in high-dimensional data without the need for a precise mathematical formulation of all physical processes. This is particularly advantageous when complete physical models are either unknown or computationally intractable. Furthermore, ML can efficiently capture multi-scale phenomena while maintaining computational efficiency during inference.

The application of machine learning models to dynamical systems spans a wide range of fields, including finance, epidemiology, rumor propagation, and predictive maintenance \citep{cheng2023machine}. Our primary motivation is the recent success of machine learning techniques in weather and climate prediction, where vast amounts of meteorological data have been analyzed to forecast weather patterns and climate changes \citep{pathak2022fourcastnet, lam2022graphcast, bodnar2024aurora}.

Machine learning models, particularly those based on deep learning, excel at processing large and complex datasets. In the context of dynamical systems, these models can be trained on historical data to learn system behavior and predict future states—essentially integrating the system over a very large time step. Among the most effective models are Recurrent Neural Networks (RNNs), Long Short-Term Memory (LSTM) networks, Gated Recurrent Units (GRUs), and Transformers \citep{chattopadhyay2020data, Vaswani2017AttentionIA}. These models are specifically designed for sequential data and retain memory of previous inputs, making them well-suited for time series prediction in dynamical systems. Convolutional Neural Networks (CNNs) have also been adapted for time series prediction, particularly in spatiotemporal problems \citep{shi2015convolutional, li2020fourier, liu2021swin, zakariaei2024advection}.

Most traditional models for dynamical systems aim for deterministic predictions, providing a single forecast of the system’s future state. However, with incomplete or noisy data, such predictions can be misleading or unrealistic due to large errors. In contrast, probabilistic approaches account for this uncertainty by forecasting a distribution of possible future states rather than a single outcome. This shift from point predictions to probability distributions is crucial for many complex systems, where small uncertainties in initial conditions or missing data can lead to vastly different outcomes, and where quantifying uncertainty is essential for decision-making.\\
{\bf Motivation:}
Machine learning techniques have made significant strides in predicting dynamical systems. However, despite their success, these models face several challenges. One of the primary difficulties lies in the quality, quantity, and completeness of the data \citep{ganesan2004coping, alanPeter, 10141545}. High-quality, large-scale datasets are essential for training accurate machine learning models, yet such data are often low resolution, incomplete, or noisy. As we show here, this violates standard machine learning assumptions. A model may receive incorrect or insufficient input (e.g. low-resolution data), preventing it from making accurate predictions. In such cases, deterministic predictions fail, necessitating the use of probabilistic forecasting, an approach that is intuitive for practitioners in the field. Instead of predicting a single outcome, one also quantifies uncertainty by leveraging historical data \citep{10141545, smith2024uncertainty}.

One way to address this issue is through ensemble forecasting \citep{kalnay2003atmospheric}. This technique is widely used in numerical weather prediction (NWP) as it accounts for uncertainties in initial conditions and model physics, producing a range of possible outcomes to enhance forecast reliability and accuracy. However, ensemble forecasting relies on extensive physical simulations and requires significant computational resources.

This study explores a framework for addressing this challenge using recent advances in machine learning. Specifically, we employ a variant of flow matching to map the current state distribution to the future state distribution. Flow matching is a family of techniques, including stochastic interpolation (SI) \citep{albergo2023stochastic, lipman2022flow} and score matching \citep{song2020sliced}, that enables learning a transformation between two distributions. While these techniques are commonly used for image generation by mapping Gaussian distributions to images, they can be readily adapted to match the distribution of the current state to the distribution of future states.

Recent work by \citet{chen2024probabilistic} introduced the use of SI and the Föllmer process for probabilistic forecasting via stochastic differential equations. Here, we propose a similar approach to handle missing and noisy data but adopt a deterministic framework instead.

Stochastic interpolation (SI) is a homotopy-based technique used in machine learning and statistics to generate new data points by probabilistically interpolating between known distributions \citep{albergo2023stochastic, lipman2022flow}. Unlike deterministic methods, which produce a single interpolated value, flow matching introduces randomness into the prediction, enabling the generation of a diverse range of possible outputs.

By leveraging probability distributions and stochastic processes, flow matching based generative models can capture the inherent variability and complexity of real-world data, resulting in more robust and versatile models.\\ 
In this work, we apply Stochastic Interpolation (SI) to probabilistic forecasting in dynamical systems and demonstrate that it is a natural choice, particularly when the dynamics are unresolved due to noise or when some states are unavailable. Specifically, we build on previous approaches that transition from Stochastic Differential Equations (SDEs) to Ordinary Differential Equations (ODEs), facilitating easier and more accurate integration \citep{yang2024consistency}.

As recently shown, selecting an appropriate loss function during training significantly improves inference efficiency \citep{lipman2024flow}. To achieve this, we further employ SI to encode and decode transport maps, introducing physical perturbations to non-Gaussian data samples. This enables sampling from the base state and propagating these samples to future states.

Finally, we demonstrate that our approach is applicable to various problems, with a key application in weather prediction, where it enhances forecasts by quantifying uncertainty.

\section{Mathematical Foundation}

In this section, we lay the mathematical foundation of the ideas behind the methods proposed in this work. Assume that there is a dynamical system of the form
\begin{eqnarray}
    \label{dynsym}
    \dot \bfy = f(\bfy, t, \bfp).
\end{eqnarray}
Here, $\bfy \in R^n$ is the state vector and $f$ is a function that depends on $\bfy$, the time $t$ and some parameters $\bfp \in R^k$.
We assume that $f$ is smooth and differentiable so that
given an initial condition $\bfy(0)$ one could compute $\bfy(T)$ using some numerical integration method \citep{stuart1998dynamical}. The learning problem arises in the case when the function $f$ or the initial conditions are unknown and we have observations about $\bfy(t)$. In this case, our goal is to learn the function $f$ given the observations $\bfy(t)$. There are a number of approaches by which this can be done \citep{reich2015probabilistic, chattopadhyay2020data, Ghadami2022}. 
Assume first that the data on $\bfy$ is measured in small intervals of time ($T=h$) that for simplicity we assume to be constant. Then, we can solve the optimization problem
\begin{eqnarray}
    \label{eq:fd}
   \min_{\bftheta}\ \hf \sum_j \left\| g\left(\hf(\bfy_j+\bfy_{j+1}), t_{j+\hf}, \bftheta \right) - {\frac {\bfy_{j+1}-\bfy_{j}}{h}} \right\|,
\end{eqnarray}
where $g$ is some functional architecture that is sufficiently expressive to approximate $f$ and $\bftheta$ are parameters in $g$. If the original function $f$ is known, then one can choose $f=g$ and $\bfp=\bftheta$. However, in many cases $g$ is only a surrogate of $f$ and the parameters $\bftheta$ are very different than $\bfp$. 

The second, less trivial case is when the time interval $T$ is very large (relative to the smoothness of the problem). In such cases, one cannot use simple finite difference to approximate the derivative. Instead, we note that since the differential equation has a unique solution and its trajectories do not intersect, we have that
\begin{eqnarray}
    \label{eq:dynsymT}
    \bfy(t+T) = F(\bfy(t), t, \bfp) = \int_t^{t+T} f(\bfy(\tau), \tau, \bfp) d\tau.
\end{eqnarray}
Here $F$ is the function that integrates the ODE from time $t$ to $t+T$. We can then approximate $F$ using an appropriate functional architecture thus predicting $\bfy$ at later time given observation on it in earlier times.

The two scenarios above fall under the case where we can predict the future based on the current values of $\bfy$. Formally, we define the following:
\begin{definition}{\bf Closed System.} \label{def1}
Let ${\cal Y}$ be the space of some observed data $\bfy$ from a dynamical system ${\cal D}$. We say the system is closed if given $\bfy(t)$ we can uniquely recover $\bfy(t+T)$ for all $t$ and finite bounded $T\le\tau$, where $\tau$ is some constant. 
Practically, given the data, $\bfy(t)$ and a constant $\epsilon$ we can estimate a function $F$ such that
\begin{eqnarray}
    \label{eq:close}
    \|\bfy(t+T) - F(\bfy(t), t, \bfp) \|^2 \le \epsilon
\end{eqnarray}
\end{definition}

Definition \ref{def1} implies that we can learn the function $F$ in \eqref{eq:dynsymT} assuming that we have sufficient amount of data in sufficient accuracy and an expressive enough architecture to approximates $F(\cdot, \cdot, \cdot)$. In this case, the focus of any ML based method should be given to the appropriate architecture that can express $F$ perhaps using some of the known structure of the problem.
This concept and its limitations are illustrated using the predator prey model. This example demonstrates how a seemingly simple dynamical system can exhibit complex behavior through the interaction of just two variables. While theoretically deterministic, the system can becomes practically unpredictable if measured in regions where the trajectories are very close.
\begin{example} {\bf The Predator Prey Model \citep[Chapter 3]{murray2007mathematical}:}
The predator prey model is
\begin{eqnarray}
    \label{eq:pp}
{\frac {d\bfy_1}{dt}} = \bfp_1\bfy_1  - \bfp_2 \bfy_1 \bfy_2 \quad {\frac {d\bfy_2}{dt}} = \bfp_3 \bfy_1 \bfy_2  -\bfp_4 \bfy_2
%    {\frac {d}{dt}} \begin{pmatrix} \bfy_1 \\ \bfy_2 \end{pmatrix} = \left\{ \begin{matrix}
%        \bfp_1\bfy_1  - \bfp_2 \bfy_1 \bfy_2 \\
%        \bfp_3 \bfy_1 \bfy_2  -\bfp_4 \bfy_2
%    \end{matrix} \right.
 \end{eqnarray}
The trajectories for different starting points are shown in \cref{fig:volot}.
\begin{figure}[t]
        \centering
        \includegraphics[width=0.8\linewidth]{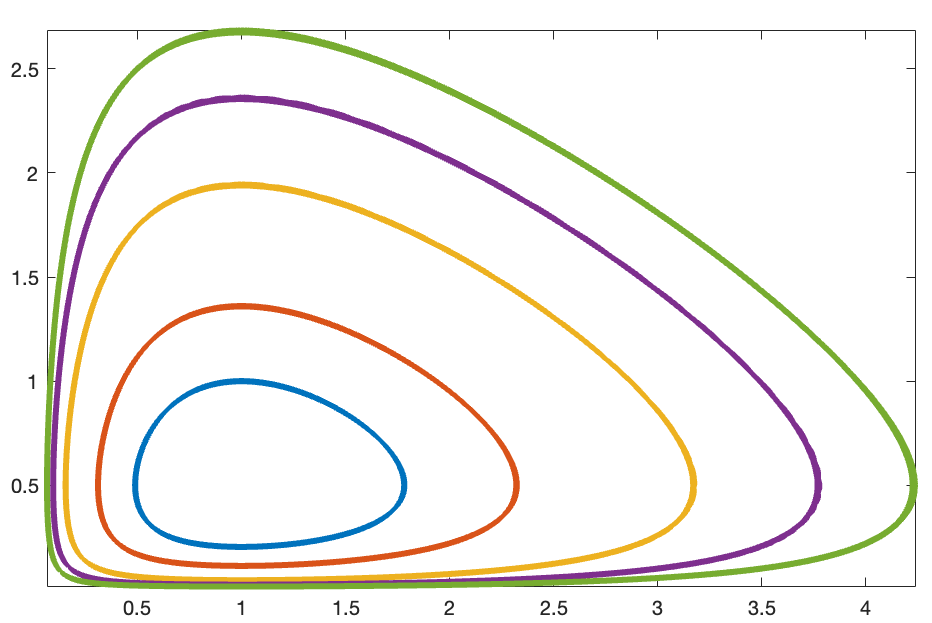}
        \caption{Trajectories for the Predator Pray Model. Note that the trajectories get very close  but do not intersect.}
        \label{fig:volot}
    \end{figure}
Assuming that we record data in sufficient accuracy, the trajectory at any time can be determined by the measurement point at earlier times, thus, justifying the model in \eqref{eq:dynsymT}. 
Furthermore, since the system is periodic, it is easy to set a learning problem that can integrate the system for a long time.

Note however, that the trajectories cluster at the lower left corner. This implies that even though the trajectories never meet in theory, they may be very close, so numerically, if the data is noisy, the corner can be thought of as a bifurcation point.

The system is therefore closed if the data $\bfy$ is accurate, however, the system is open if there is noise on the data such that late times can be significantly influenced by earlier times.
\end{example}

The predator prey model draws our attention to a  different scenario that is much more common in practice. Assume that some data is missing or, that the data is polluted by noise. For example, in weather prediction, where the primitive equations are integrated to approximate global atmospheric flow, it is common that we do not observe the pressure, temperature or wind speed in sufficient resolution. In this case, having information about the past is, in general, insufficient to predict the future.
In this case we define the system as open.
%\begin{definition} \label{def2}{\bf Open System.}
%Let ${\cal Y}$ be the space of some observed data $\bfy$ on a dynamical system ${\cal D}$. We say the system is open if given $\bfy(t)$ we {\bf cannot} uniquely recover $\bfy(t+T)$ for all $t$  and finite $T\le\tau$, where $\tau$ is some constant. That is, there is a constant $\epsilon$ such that
%\begin{eqnarray}
%\label{eq:open}
%    \|\bfy(t+T) - F(\bfy(t), t, \bfp) \| \ge \epsilon
%\end{eqnarray}
%independent of the amount of data and the %complexity of $F$.
%\end{definition}

While we could use classical machine learning techniques to solve closed systems it is unreasonable to attempt and solve open systems with similar techniques. This is because the system does not have a unique solution, given partial or noisy data. To demonstrate that we return to the predator prey model, but this time with partial or noisy data.

\begin{example} \label{ex2} {\bf The predator Pray Model with Partial or Noisy Data:}
Consider again the predator prey model, but this time assume that we record $\bfy_1$ only. 
Now assume that we known that at $t=0$, $\bfy_1=1$ and our goal is to predict the solution at time $t=200$, which is very far into the future.  Clearly, this is impossible to do. Nonetheless, we can run many simulations where $\bfy_1 = 1$ and $\bfy_2 = \pi(\bfy_2)$ where $\pi$ is some density. For example, we choose $\pi(\bfy_2) = U(0,1)$. In this case we ran simulations to obtain the results presented in \cref{fig:ts} (left). Clearly, the results are far from unique. Nonetheless, the points at $t=200$ can be thought of as samples from a probability density function. This probability density function resides on a curve. Our goal is now transformed from learning the function $F(\bfy,t)$ in~\eqref{eq:dynsymT} to learn a probability density function
$\pi(\bfy(T))$.
\begin{figure}[t]
        \centering
        \begin{tabular}{cc}
        \includegraphics[width=0.45\linewidth]{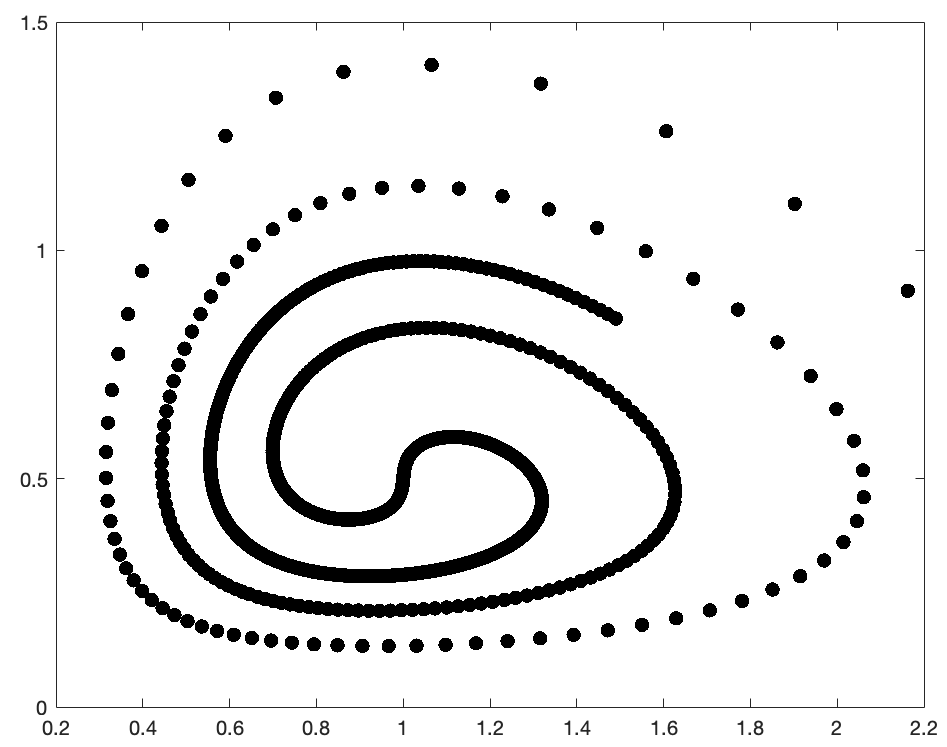} &
        \includegraphics[width=0.44\linewidth]{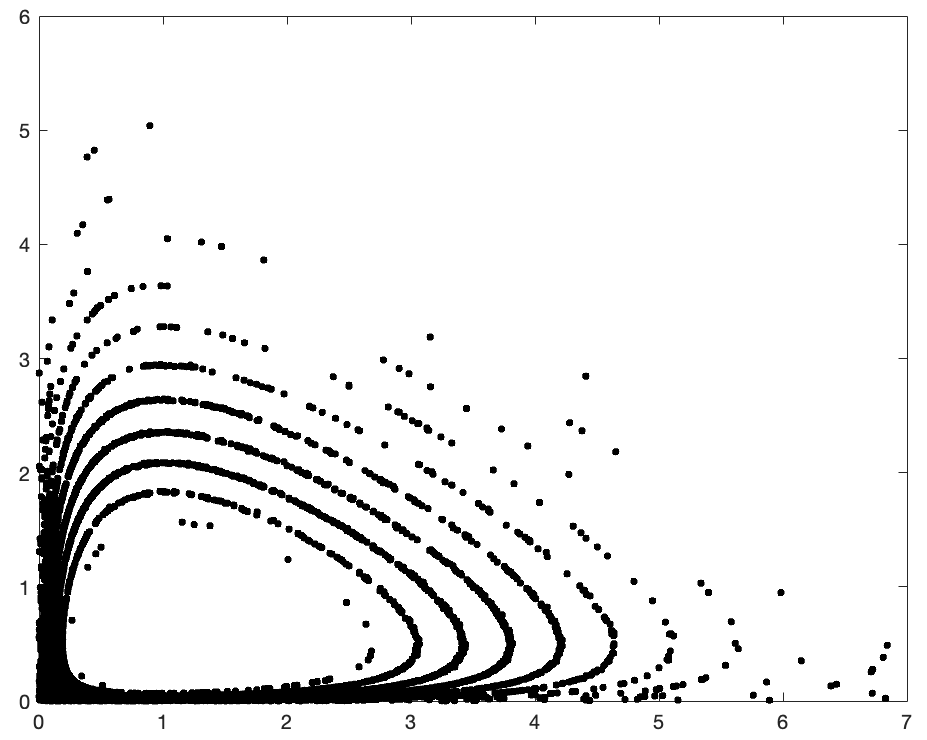}
        \end{tabular}
        \caption{Left: The solution for $\bfy_1(0)=1$ and $\bfy_2(0) \sim U(0,1)$ at $t=200$. Right:The solution for $\bfy(0)=[0.1, 0.3]^{\top} + \epsilon$ where $\epsilon \sim N(0, 0.05 \bfI)$ at $t=200$}
        \label{fig:ts}
\end{figure}

A second case that is of interest is when the data is noisy. Consider the case that $\bfy_0^{\rm obs} = \bfy(0) + \epsilon$. In this case we consider $\bfy_0 = [0.1,0.3]^{\top}$ and we contaminate it with Gaussian noise with $0$ mean and $0.05$ standard deviation. 
Again, we attempt to predict the data at $T=200$. 
We plot the results from 1000 simulations in \cref{fig:ts} (right). Again, we see that the incomplete information transforms into a density estimation problem.
\end{example}

Example~\ref{ex2} represents a much more realistic scenario, since in reality data is almost always sampled only partially or on low resolution and includes noise. 

\begin{comment}
    It is important to note that in many cases one can transform an open system into a closed one by considering higher order dynamics. For example, in the preditor prey model, one could (formaly) eliminate $\bfy_2$ by setting
    
    $$ \bfy_2 = (\bfp_2 \bfy_1)^{-1}\left(\dot \bfy_1 - \bfp_1 \bfy_1 \right) $$
    
    and then subsituting $\bfy_2$ into the second system, obtaining a second order system. This however, requires data that is sufficiently accurate to be differentiated more that once.
\end{comment}
Motivated by the two examples above, we now form the learning problem of probabilistic forecasting (see, e.g. \cite{reich2015probabilistic}). 
\begin{definition}
    \label{def:stoch}{\bf Probabilistic forecasting.}
    Probabilistic forecasting predicts future outcomes by providing a range of possible scenarios along with their associated probabilities, rather than a single point estimate. \\
    To be specific, let the initial vector $\bfy(0) \sim \pi_0(\bfy)$ and assume that $\bfy(T)$ is obtained from integrating the dynamical system ${\cal D}$. Probabilistic forecasting refers to estimating and sampling from the distribution $\pi_T(\bfy)$.  
\end{definition}

This approach acknowledges that a unique prediction may not be attainable but allows for generating samples of future outcomes. In contrast, deterministic prediction assumes a closed system and highly accurate data, making it often inapplicable. 
%In the next section we discuss how to achieve probabilistic forecasts using stochastic interpolation.

\section{Probabilistic Forecasting and Stochastic Interpolation}
\label{sec:probForc}

In this section, we discuss a version of Stochastic Interpolation (SI). SI is the technique that we use to propagate the uncertainty and also sample from the initial distribution.

To this end,
let us consider the case that we observe a noisy or a part of the system. In particular, let $\bfy(t)$ be  a vector that is obtained from an unknown dynamical system ${\cal D}$ and let
\begin{eqnarray}
    \bfq_0(t) = \bfS \bfy(t) + \bfepsilon
\end{eqnarray}
Here $\bfS$ is some sampling operator that typically reduces the dimension of $\bfy$ and $\bfepsilon \sim N(0, \sigma^2\bfI)$.
The distinction between $\bfy$ and $\bfq_0$ is important. The vector $\bfy$ represents the full state of the system, while $\bfq_0$ represents only partial knowledge of the system. For many systems, it is impossible to observe the complete
state space. For example, for local weather prediction we may have temperature stations but not wind stations, where wind speed is essential to model the advection of heat. The inability to obtain information about the full state implies that most likely it is impossible to deliver a deterministic prediction, and thus stochastic prediction is the natural choice. 

Given $\bfq_0(t)$,
our goal is to sample from the distribution of observed partial state $\bfq_T = \bfq(t+T) \sim \pi_T(\bfq)$,  given samples from the distribution on the data at $\bfq_0(t) \sim \pi_0(\bfq)$.
Clearly, we do not have a functional expression for neither $\pi_0$ nor $\pi_T$, however, if we are able to obtain many samples from both distributions, we can use them to build such a distribution. This assumption is unrealistic without further assumptions in the general case  since we typically have only a single time series.
A few assumptions are common and we use
them in our experiments. The most common assumptions are, i)
 The time dependent process is autoregressive. This implies that we can treat every instance (or instances) of the time series as $\bfq_0$ and use the future time series as $\bfq_T$. Such an assumption is very common although it may be unrealistic, and ii)
Periodicity and clustering. For many problems, especially problems that relate to earth science, there is a natural periodicity and data can be clustered using the natural cycle of the year. 

Given our ability to obtain data that represents samples from the probability at time $0$ and at time $T$, we are able to use
the data and provide a stochastic prediction.
To this end, we note that stochastic prediction is nothing but a problem of probability transformation which is in the base of mass transport \citep{BB2000}. 
While efficient solutions for low dimensions problems have been addressed \citep{fisher1970statistical, villani2009optimal}, solutions for problems in high dimensions are only recently being developed.

One recent and highly effective technique to solve such a problem is stochastic interpolation (SI) \citep{albergo2022building}, which is a technique that belongs to a family of flow matching methods \citep{lipman2022flow,albergo2023stochastic,song2020sliced}. The basic idea of stochastic interpolation
is to generate a simple interpolant for all possible
points in $\bfq_0$ and $\bfq_T$. A simple interpolant of this kind is linear and reads
\begin{eqnarray}
    \bfq_t = t\bfq_T + (1-t)\bfq_0
\end{eqnarray}
where $\bfq_T \sim \pi_T$, $\bfq_0 \sim \pi_0$, and $t\in[0,1]$ is a parameter. The points $\bfq_t$ are associated from a distribution $\pi_t(\bfq)$ that converges to $\pi_T(\bfq)$ at $t=1$ and to $\pi_0(\bfq)$ at $t=0$.
In SI one learns (estimates) the velocity 
\begin{eqnarray}
   \bfv_t(\bfq_t) = \dot\bfq_t =\bfq_T -\bfq_0 
\end{eqnarray}
by solving the stochastic optimization problem
\begin{eqnarray}
    \label{eq:vel_learn}
    \min_{\bftheta} {\frac 12}{\mathbb E}_{\bfq_0, \bfq_1}\int_0^1 \|\bfv_{\bftheta}(\bfq_t, t) - (\bfq_T -\bfq_0) \|^2 \, dt
\end{eqnarray}
Here $\bfv_{\bftheta}(\bfq_t, t)$ is an interpolant for the velocity $\bfv$ that is given at points $\bfq_t$. A common model that is used for $\bfv_{\bftheta}(\bfq_t, t)$ is a deep neural network. While for simple problems such models can be composed of a simple network with a small number of hidden layers, for complex problems, especially problems that relate to space-time predictions, complex models such as U-Nets are commonly used \citep{ronneberger2015u}. 

Assume that the a velocity model $\bfv_{\bftheta}(\bfq_t, t)$ is trained
and uses it to integrate $\bfq$ from time $0$ to $T$, that is
\begin{eqnarray}
    \label{eq:ode}
    {\frac {d\bfq}{dt}} = \bfv_{\bftheta}(\bfq, t) \quad \bfq(0) = \bfq_0
\end{eqnarray}
Note that we use a deterministic framework rather than a stochastic framework. This allows to incorporate high accuracy integrators that can take larger step sizes.
However, this implies that the solution of the ODE given a single initial condition $\bfq_0$ gives a single prediction. In order to sample from the target distribution, $\pi_T$ we sample from $\bfq_0$ and then use the ODE (\eqref{eq:ode}) to push many samples forward. 
We thus obtain $M$ different samples for $\bfq_0$ (see next section) and use them in order to sample from $\pi_T$.
\begin{comment}
    Note that the ODE obtained for $\bfq$ is not physical. It merely used to interpolate the density from time $0$ to $T$. To demonstrate we continue with the predator prey model.
\end{comment}

\section{Sampling the Perturbed States} \label{sec:GeneratePerturbation}

Sampling from the distribution $\bfq_0$ in a meaningful way is not trivial. There are a number of approaches to achieve this goal.
A brute force approach can search through the data for so-called similar states, for example, given a particular state $\bfq_0$ we can search other states in the data set such that
$\|\bfq_0 - \bfq_i\|^2 \le \epsilon$. While this approach is possible in low dimension it is difficult if not impossible in very high dimensions. For example, for global predictions, it is difficult to find two states that are very close everywhere.
To this end, we turn to a machine learning approach that is designed to generate realistic perturbations.

We turn our attention to Variational Auto Encoders (VAE's). In particular, we use a flow based VAE \citep{dai2019diagnosing, vahdat2021score}. 
Variational Autoencoders are particularly useful. In the encoding stage, they push the data $\bfq_0$ from the original distribution $\pi_0(\bfq)$ to a vector $\bfz$ that is sampled from a Gaussian, $N(0,\bfI)$. Next, in the decoding stage, the model pushes the vector $\bfz$ back to $\bfq_0$. Since the Gaussian space is convex, small perturbation can be applied to the latent vector $\bfz$ which lead to samples from $\pi_0(\bfq)$ that are centered around $\bfq_0$.

The difference between standard VAEs and flow based VAEs is that give $\bfq_0 \sim \pi_0$, VAEs attempt to learn a transformation to a Gaussian space $N(0,\bfI)$ directly. However, as have been demonstrated in \cite{ruthotto2021introduction} VAEs are only able to create maps to a latent state that is similar to but not Gaussian, making it difficult to perturb and sample from them.
VAEs that are flow based can generate a Gaussian map to a much higher accuracy. Furthermore, such flows can learn the decoding from a Gaussian back to the original distribution.  Flows that can transform the points between the two distributions are sometimes referred to as symmetric flows \citep{ho2020denoising, albergo2023stochastic}.  

Such flows can be considered as a special case of SI, an encoder-decoder scheme for a physical state $\bfq_0$ being encoded to a Gaussian state $\bfz$. 
To this end, let us define the linear interpolant  as
\begin{eqnarray}
\bfq_t =
\begin{cases}
    (1 - 2t)\bfq_0 + 2t\bfz,& \text{if } t \in [0, \hf)\\
    (2t - 1)\bfq_0 + 2(1-t)\bfz,              & \text{if } t \in [\hf, 1]
\end{cases}
\end{eqnarray}
The velocities associated with the interpolant as simply
\begin{eqnarray}
\bfu_t = {\frac {d\bfq_t}{dt}} =
\begin{cases}
    2(-\bfq_0 + \bfz) ,& \text{if } t \in [0, \hf)\\
    2(\bfq_0 - \bfz),              & \text{if } t \in [\hf, 1]
\end{cases}
\end{eqnarray}
Note that the flow starts at $\bfq_0 \sim \pi_0$ towards $\bfz \sim N(0,\bfI)$ at $t=0$ and arrive to $\bfz$ at $t=\hf$. This is the encoding state. In the second part of the flow we learn a decoding map that pushes the points $\bfz$ back to $\bfq_0$.
Note also that $\bfu$ is symmetric about $t=1/2$
which is used in training.
%$$ \bfu_t(\bfq_t, t) = -\bfu_t(\bfq_t, 1-t) \quad 0<t<\hf $$
Training these models is straight forward and is done
in a similar way to the training of our stochastic interpolation model, namely we solve a stochastic optimization problem of the form
\begin{eqnarray}
    \label{eq:vel_learnz}
    \min_{\bftheta} {\frac 12}{\mathbb E}_{\bfq_0, \bfz}\int_0^{\hf} \|\bfu_{\bftheta}(\bfq_t, t) + 2(\bfq_0 - \bfz) \|^2 dt
\end{eqnarray}
Note that we can use the same velocity for the reverse process. Given the velocity $\bfu_{\bftheta}$ we now generate perturb samples in the following way. First, we integrate the ODE to $t=\frac 12$, that is we solve the ODE
\begin{eqnarray}
    \label{eq:ode1}
    {\frac {d\bfq}{dt}} = \bfu_{\bftheta}(\bfq, t) \quad \bfq(0) = \bfq_0 \quad t \in [0, \frac 12]
\end{eqnarray}
This yield the state $\bfq(\frac 12) \sim N(0, \bfI)$.
We then perturb the state 
\begin{eqnarray}
    \label{eq:pert}
    \widehat \bfq(1/2) = \bfq(1/2) + \sigma \boldsymbol{\omega}
\end{eqnarray}
where $\boldsymbol{\omega} \sim N(0,\bfI)$ and $\sigma$ is a hyper-parameter. We then integrate the ODE \eqref{eq:ode1} from $\hf$ to $1$ starting from $\widehat \bfq(\hf)$ obtaining a perturb state $\widehat \bfq_0$. The integration is done in batch mode, that is, we integrate $M$ vectors simultaneously to obtain $M$ samples from the initial state around $\bfq_0$. These states are then used to obtain $M$ samples from $\bfq_T$ as explained in \cref{sec:probForc}.

%\textcolor{red}{Sid, we need a few pictures that show the perturbations that are generated from this approach}

\section{Experiments}\label{sec:expt}

Our goal in this section is to apply the proposed framework to a set of various temporal datasets and show how it can be used to estimate the uncertainty in a forecast. We experiment with two synthetic datasets that can be fully analyzed and a realistic dataset for weather prediction. Further experiments on other data sets can be found in the appendix.

\subsection{Datasets}\label{subsec:datasets}
We now describe the datasets considered in our experiments.
\textbf{Lotka–Volterra predator–prey model}: This is a nonlinear dynamical system. It follows the equation \ref{eq:pp}, where $\bfp_0$ = 2/3, $\bfp_1$ = 4/3, $\bfp_3$ = 1 and $\bfp_4$ = 1. The initial distribution of states that is Gaussian with a mean of $\bfy(0) = [0.1, 0.3]^T$ and standard deviation of $0.05$ the final distribution of states is a distribution obtained after fine numerical integration over $t \in [0, 200]$. Note that the length of the integration time is very long which implies that the output probability space is widely spread which makes this simple problem of predicting the final distribution from an initial distribution very difficult. 

\textbf{MovingMNIST}: Moving MNIST is a variation of the well known MNIST dataset \citep{MovingMNIST}. The dataset is a synthetic video dataset designed to test sequence prediction models. It features 20-frame sequences where two MNIST digits move with random trajectories. Following the setup described in \cite{MovingMNIST}, we trained our model on 10,000 randomly generated trajectories and then used the standard publicly available dataset of 10,000 trajectories for testing.

\textbf{WeatherBench}: The dataset used for global weather prediction in this work is derived from WeatherBench \citep{weatherBench}, which provides three scaled-down subsets of the original ERA5 reanalysis data \citep{era5hersbach}. The specific variables and model configurations are detailed in \cref{app:WBvars}. Our forecasting models use the 6-hour reanalysis window as the model time step.

The statistics and configurations for each of the datasets during experiments are mentioned in the appendix \ref{app:datasetstats}. We have two additional real datasets, Vancouver93 and CloudCast, in the appendix for additional analysis. In the next section we give details for experiments and results on the predator-prey model, the MovingMNIST and the WeatherBench datasets. 

\subsection{Results}
Our goal is to learn a homotopy based deterministic functional map for complex timeseries to predict the future state from the current state but with uncertainty. To do so, we evaluate an ensemble of predictions from an ensemble of initial states. The initial states are obtained by using the auto-encoder described 
in Section~\ref{sec:GeneratePerturbation}. We then push those states forward obtaining an ensemble of final states. We then report on the statistics of the final states. 

\begin{figure}[t]
    \centering
\includegraphics[width=1.0\linewidth]{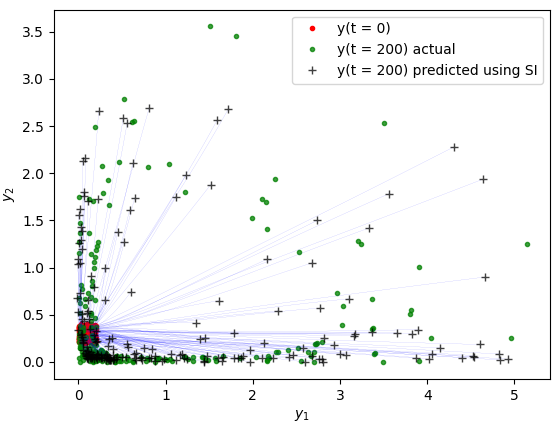}
    \caption{Comparison of actual final distribution and that obtained using SI on the predator-prey Model. Trajectories for transport learned by SI are in blue. Note that the trajectories are not physical.}
    \label{fig:pred-prey200}
\end{figure}

{\bf The Predator Pray Model:}
We apply SI to learn a deterministic functional map on predator-prey model  \ref{subsec:datasets} for a long time integration. The initial state of the data is sampled as a group of noisy initial states, that is, we consider the case that 
$\bfq = \bfy(0) + \epsilon. $ 
In this case we consider $\bfy(0) = [0.1,0.3]^{\top}$ and we contaminate it with the noise vector, $\epsilon \sim N(0,0.05\bfI)$. Our goal is to predict the state, $\bfq(T)$ at time $T=200$. The distribution of 250 noisy points for $\bfq(0)$ and the final distributions for $\bfq(200)$ using numerical integration and SI can be seen in \cref{fig:pred-prey200}. For this simple case, a multilayer perceptron (MLP) is used to learn the mapping. It can be noticed that SI is able to match with a sample from the final complex distribution, specifically the areas with high probabilities are captured very well. Some more statistical comparisons of the variables' state are reported in appendix \ref{app:predpreystats}, to support our results. Appendix \ref{Van93results} shows some results on Vancouver93 where the learned final distribution from simple distribution to a complex conditional distribution is clearly evident. 

The same concept can be extended to high-dimensional spatiotemporal image space instead of vector space without any loss of generality. Like in \cite{lipman2022flow, Karras_2024_CVPR,bieder24a}, we use a U-Net architecture from \cite{dhariwal2021diffusion} to learn the velocity.  

\begin{figure*}[!h]
    \centering
    \includegraphics[width=1.0\linewidth]{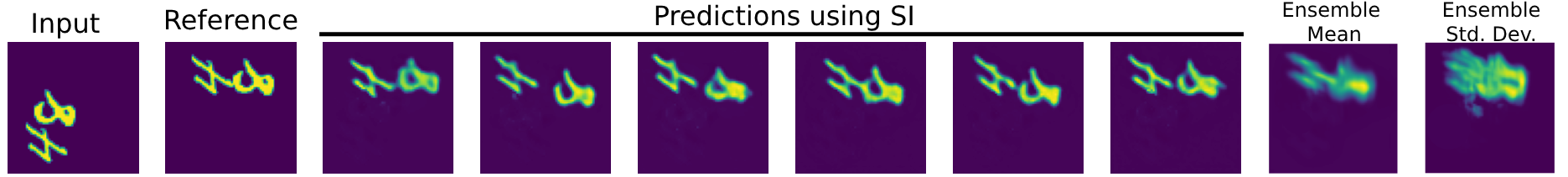}
    \caption{Six of 50 Moving MNIST trajectory predictions obtained using SI and their ensemble mean and standard deviation.}
    \label{fig:MMMultiPred}
\end{figure*}

{\bf Moving MNIST:}
We use the Moving MNIST to train a network (U-Net) to predict 10 time frames into the future, given 10 past frames.  The hyperparameter settings of the U-Net are in appendix \ref{UNetsettings}. A set of initial states is obtained using the auto-encoder creating 50 initial and final states. Using SI and the algorithm proposed in \ref{sec:probForc} we sample predictions of final states. 
The predictions can be observed in \cref{fig:MMMultiPred}. The predictions are similar yet different from the true final state.
Statistical comparisons of the ensemble predictions using SI with the ensemble of real targets can be seen in \cref{Result:genMetrics} and in \cref{app:MMstats}.  

\begin{figure*}[!h]
    \centering
    \includegraphics[width=1.0\linewidth]{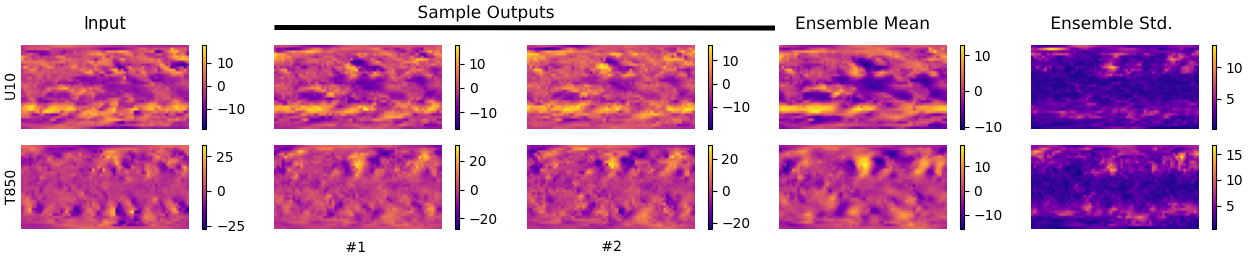}
    \caption{Two sample stochastic forecasts of U10 and T850 after 2 days obtained using SI and the ensemble mean and standard deviation for 78 forecasts.}
    \label{fig:wBsampleforecasts}
\end{figure*}

\paragraph{ WeatherBench:}
We now use the WeatherBench dataset to build a forecasting model using SI. We use a similar U-Net as in Moving MNIST for learning the functional mapping. The hyperparameter settings are in appendix \ref{UNetsettingswB}. A particular time step is selected from the testset and 77 most similar samples are searched from the testset to collect the initial states, where similarity in measured by the $L_2$ norm.
We hence have the ensemble of actual initial and final states of the true data. This ensample is used as ground truth and allows an unbiased testing of our method. Using our technique, we generate an ensemble of initial and final states. A few samples of generated states along with the mean and standard deviation of the ensemble of 78 forecasts can be seen in \cref{fig:wBsampleforecasts}. We use the standard variables U10 and T850 for visual observation here. Some statistical comparisons of our ensemble prediction, variables Z500, T2m, U10 and T850, can be seen in \cref{Result:genMetrics} and in \cref{app:wBstats}. 

\paragraph{Metrics:} \label{Result:genMetrics}
Below, we describe the metrics used to evaluate the performance of our model across different datasets. For each ensemble of predictions generated by our model, there is a corresponding ensemble of target states. The goal is to ensure that the statistical characteristics of the predicted ensemble closely match those of the target ensemble.
The simplest metric for comparing these high-dimensional distributions is the \textbf{mean score}, which represents the average of the pixel values in a state. Similarly, we compute the \textbf{standard deviation score}. For two distributions to be considered similar, these scores should ideally align. More similar the scores are two distributions, more similar are the distributions. Table \ref{tab:statsMetrics} presents the mean and standard deviation scores for both the target ensemble and predicted ensemble, demonstrating a strong alignment between the two. 
Inspired by methods used in turbulence flow field analysis, we further compare the predicted and target ensembles by computing their mean state and standard deviation state. To quantify the similarity between these states, we employ standard image comparison metrics: Mean Squared Error (MSE), Mean Absolute Error (MAE), and Structural Similarity Index Measure (SSIM). These metrics are applied to both the ensemble mean states and the ensemble standard deviation states, as shown in Tables \ref{tab:similarityMetricsMean} and \ref{tab:similarityMetricsStd}.
Lower the MSE and MAE are, better the uncertainty is captured. Higher the SSIM, the uncertainty is captured. 

As can be readily observed in the tables, our approach provides an accurate representation of the mean and standard deviation, which are the first and second order statistics of each data set. The definition of the metrics are in \cref{app:metrics}.

\begin{table}[t]
\centering
\begin{tabular}{lcccc}
\toprule
 & {True}& {Our} & {True}& {Our}  \\
 & {Mean}& {Mean} & {Std Dev}& {Std Dev}  \\
{Data}                 & {Score}& {Score} & {Score}& {Score}  \\\midrule
Predator–Prey          & 7.55e-1 & 7.43e-1     & 1.04  & 1.07        \\
MovingMNIST            & 6.01e-2 & 5.58e-2     & 2.21e-1  & 1.87e-1         \\
WeatherBench            & 1.36e4 & 1.36e4     & 9.15e1  & 9.66e1         \\
\bottomrule
\end{tabular}
\caption{Comparison of mean score and standard deviation score of ensemble predictions.}
\label{tab:statsMetrics}
\end{table}

\begin{table}[t]
\centering
\begin{tabular}{lcccc}
\toprule
 {Data}                & {MSE($\downarrow$)}& {MAE($\downarrow$)} & {SSIM($\uparrow$)}\\\midrule
Predator–Prey          & 8.0e-3 & 8.8e-2          & NA  \\
MovingMNIST            & 2.5e-3 & 1.8e-2          & 0.843  \\
WeatherBench           & 1.9e4  & 5.2e1           & 0.868  \\
\bottomrule
\end{tabular}
\caption{Accuarcy of our ensemble mean.}
\label{tab:similarityMetricsMean}
\end{table}

\begin{table}[t]
\centering
\begin{tabular}{lcccc}
\toprule
 {Data}                & {MSE($\downarrow$)}& {MAE($\downarrow$)} & {SSIM($\uparrow$)}\\\midrule
Predator–Prey          & 2.7e-2 & 1.6e-1 & NA  \\
MovingMNIST            & 3.6e-3 & 2.5e-2 & 0.744  \\
WeatherBench           & 1.1e4  & 3.6e1  & 0.608  \\
\bottomrule
\end{tabular}
\caption{Accuracy of our ensemble standard deviation.}
\label{tab:similarityMetricsStd}
\end{table}

\begin{figure}[!h]
    \centering
    \includegraphics[width=\linewidth]{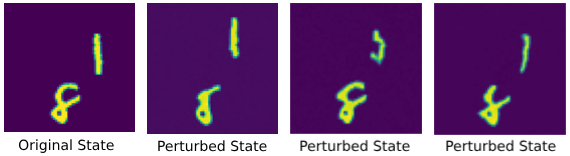}
    \caption{Three random perturbed states of a Moving MNIST sample state.}
    \label{fig:MMperturbed}
\end{figure}

\begin{figure}[!h]
    \centering
    \includegraphics[width=\linewidth]{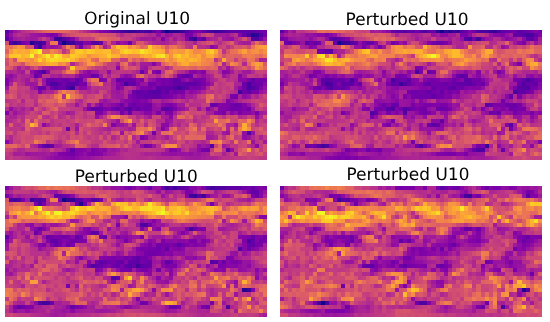}
    \caption{Three random perturbed states of a WeatherBench U10 state.}
    \label{fig:wBperturbedU10}
\end{figure}

\begin{figure}[!h]
    \centering
    \includegraphics[width=\linewidth]{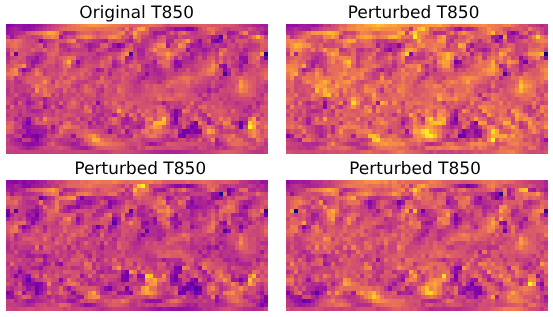}
    \caption{Perturbed states of a WeatherBench T850 state.}
    \label{fig:wBperturbedT850}
\end{figure}

\paragraph{Perturbation of Physical States using SI:}
\label{exPhysicalPerturbation}
Perturbation of a physical state in a dynamical system refers to a deliberate natural deviation in a system's parameters or state variables. Using the algorithm proposed in \ref{sec:GeneratePerturbation}, we generate perturbed states for MovingMNIST, and WeatherBench. All the generated perturbed states are very much realistic. Some sample perturbed states from MovingMNIST can be seen in \cref{fig:MMperturbed}. The physical disturbances like bending or deformation of digits, change in position and shape is clearly captured. Similarly, we generated perturbed states for WeatherBench as well. Some samples of perturbed states from WeatherBench's U10 and T850 variables (as these are easier to visualise for humans) can be seen in \cref{fig:wBperturbedU10} and \cref{fig:wBperturbedT850} respectively. As it can be seen the perturbations are very much physical. Additional studies can be seen in \cref{app:MM}.    

\section{Conclusion}\label{sec:conclusion}

Probabilist forecasting is an important topic for many scientific problems. It expands traditional machine learning techniques beyond paired input and outputs, to mapping between distributions. 
While generative AI has been focused on mapping Gaussians to data, similar methodologies can be applied for the generation of predictions given current states. 
In this work we have coupled a stochastic interpolation to propagate the current state distribution with an auto-encoder that allows us to sample from the current distribution. We have shown that this approach can lead to a computationally efficient way to sample future states even for long integration times and for highly non-Gaussian distributions in high dimensions.

% References
\bibliography{arXiv2025-main}

\newpage

\onecolumn

\title{Appendix}
\maketitle 
\appendix

\section{Additional Datasets} \label{app:addDatasets}
\subsection{Datasets} 
\begin{itemize}
    \item \textbf{Vancouver 93 Temperature Trend (Vancouver93)}: This is a real nonlinear chaotic high dimensional dynamical system, in which only a single state (temperature) is recorded.  The daily average temperatures at 93 different weather stations in and around Vancouver, BC, Canada are captured for the past 34 years from sources such as National Oceanic and Atmospheric Administration (NOAA), the Government of Canada, and Germany's national meteorological service (DWD) through Meteostat's python library \cite{lamprechtmeteostat}.  Essentially, it is a time series of 12,419 records at 93 stations. The complex urban coastal area of Vancouver with coasts, mountains, valleys, and islands makes it an interesting location where forecasting is very difficult than in general \cite{vannini2012making, leroyer2014subkilometer}. The historical temperature alone insufficient to predict the temperature in the future, as it requires additional variables like precipitation, pressure, and many more variables at higher resolution.
This makes the dataset fit the proposed framework.
  \item \textbf{CloudCast} \cite{nielsen2020cloudcast}: a real physical nonlinear chaotic spatiotemporal dynamical system. The dataset comprises 70,080 satellite images capturing 11 different cloud types for multiple layers of the atmosphere annotated on a pixel level every 15 minutes from January 2017 to December 2018 and has a resolution of 128$\times$128 pixels (15$\times$15 km).
\end{itemize}

\section{Datasets} \label{app:datasetstats}
Table \ref{tab:datasetstats} describes the statistics, train-test splits, image sequences used for modeling, and frame resolutions.

\begin{table}[!h]
\centering
\vspace{10pt}
\begin{tabular}{lccccc}
\toprule
{Dataset} & \( N_{\text{train}} \) & \( N_{\text{test}} \) & \( (C, H, W) \) & History & Prediction \\
\midrule
Predator-Prey           & 10,000                  & 500                & (1, 1, 2)     &    1   & 1       \\
Vancouver93            & 9,935                  & 2,484                & (93, 1, 1)     & 5 (5 days)      & 5 (5 days)       \\
Moving MNIST           & 10,000                  & 10,000                & (1, 64, 64)     & 10      & 10       \\
CloudCast        & 52,560                  & 17,520                 & (1, 128, 128)   & 4 (1 Hr)       & 4 (1 Hr)       \\
WeatherBench (5.625$^{\circ}$)        & 324,336                  & 17,520                 & (48, 32, 64)   & 8 (48 Hrs)       & 8 (48 Hrs)       \\
\bottomrule
\end{tabular}
\caption{Datasets statistics, training and testing splits, image sequences, and resolutions}
\label{tab:datasetstats}
\end{table}

\subsection{WeatherBench} \label{app:WBvars}
 The original ERA5 dataset \cite{era5hersbach} has a resolution of 721$\times$1440 recorded every hour over almost 40 years from 1979 to 2018 across 37 vertical levels for 0.25$^\circ$ latitude-longitude grid. The raw data is extremely bulky for running experiments, even with powerful computing resources. We hence the typically used reduced resolutions (32$\times$64: 5.625$^\circ$ latitude-longitude grid, 128$\times$256: 5.625$^\circ$ latitude-longitude grid) as per the transformations made by  \cite{weatherBench}. We, however, stick to using the configuration set by \cite{nguyen2023climax} for standard comparison. The prediction model considers 6 atmospheric variables at 7 vertical levels, 3 surface variables, and 3 constant fields, resulting in 48 input channels in total for predicting four target variables that are considered for most
medium-range NWP models like the state of the art IFS\cite{IFSwedi2015modelling} and are often used for benchmarking in previous deep learning work as well like \cite{graphcast2023,fourcastnet2022pathak}, they are geopotential at 500hPa (Z500), the temperature at 850hPa (T850), the temperature at 2 meters from the ground (T2m), and zonal wind speed at 10 meters from the ground (U10). We use a leap period of 6 hours as a single timestep and hence our model takes in 8 timesteps (48 hours or 2 days) to predict for the next 8 timesteps (48 hours or 2 days). According to the same setting used the data from 1979 to 2015 as training, for the year 2016 as validation set and for the years 2017 and 2018 as testing set. The details of the variables  considered are in Table \ref{tab:wB_config}.

\begin{table}[!h]
\centering
\begin{tabular}{clccc}
\hline
Type & Variable Name & Abbrev. & ECMWF ID  & Levels \\
\hline
 & Land-sea mask & LSM & 172 \\
Static & Orography & OROG & 228002\\
 & Soil Type & SLT & 43\\
\hline
 & 2 metre temperature & T2m & 167 \\
Single & 10 metre U wind component & U10 & 165 \\
 & 10 metre V wind component & V10 & 166 \\
\hline
 & Geopotential & Z & 129 & 50, 250, 500, 600, 700, 850, 925 \\
 & U wind component & U & 131 & 50, 250, 500, 600, 700, 850, 925 \\
Atmospheric & V wind component & V & 132 & 50, 250, 500, 600, 700, 850, 925 \\
 & Temperature & T & 130 & 50, 250, 500, 600, 700, 850, 925 \\
 & Specific humidity & Q & 133 & 50, 250, 500, 600, 700, 850, 925 \\
 & Relative humidity & R & 157 & 50, 250, 500, 600, 700, 850, 925 \\
\hline
\end{tabular}
\caption{Variables considered for global weather prediction model.}
\label{tab:wB_config}
\end{table} 

\section{Comparison Metrics} \label{app:metrics} 
\subsection{Ensemble Mean and Ensemble Standard Deviation}  
Let $S = {I_1,..,I_N}$ be a set of images, $I_i \in \cal{R}^{C\times H \times W}$ such that $i,j,k,N,C,H,W \in \mathbb{N}$, $i\leq N$, $j\leq C$, $k\leq H$, and $ l \leq W$. 
An image $I_i$ is defined as 
\[
I_i = \{ P_i^{j,k,l} \in \cal{R} | \text{j} \leq C, \text{k} \leq H, \text{l} \leq W\}, 
\]
where $P_i^{j,k,l}$ is called a pixel in an image $I_i$.

Ensemble mean state (image), $I_{EM} \in \cal{R}^{C\times H \times W }$, is defined as

\begin{equation}
I_{EM} = \{ P_{EM}^{j,k,l} = \frac{1}{N} \sum_{i=1}^{N} P_i^{j,k,l} | P_i^{j,k,l} \in I_i\}.
\end{equation}

Ensemble mean score $V_{EM}$ is defined as
\begin{equation}
V_{EM} = \frac{1}{N\cdot C \cdot H \cdot W} \sum_{i=1}^{N} \sum_{j=1}^{C} \sum_{k=1}^{H} \sum_{l=1}^{W} P_i^{j,k,l},
\end{equation}
where $ P_i^{j,k,l} \in I_i$.

Ensemble standard deviation state (image), $I_{ES} \in \cal{R}^{C\times H \times W }$, is defined as

\begin{equation}
I_{ES} = \{ P_{ES}^{j,k,l} = \sqrt{\frac{1}{N} \sum_{i=1}^{N} (P_i^{j,k,l} - P_{EM}^{j,k,l})^2} | P_i^{j,k,l} \in I_i, P_{EM}^{j,k,l} \in I_{EM}\}.
\end{equation}

Ensemble variance score $V_{ES}$ is defined as
\begin{equation}
V_{ES} = \frac{1}{N\cdot C \cdot H \cdot W} \sum_{i=1}^{N} \sum_{j=1}^{C} \sum_{k=1}^{H} \sum_{l=1}^{W} (P_i^{j,k,l} - V_{EM})^2,
\end{equation}
where $ P_i^{j,k,l} \in I_i$.

\subsection{MSE, MAE, SSIM}

\begin{equation}
\text{MSE} = \frac{1}{N\cdot C \cdot H \cdot W} \sum_{i=1}^{N} \sum_{h=1}^{H} \sum_{w=1}^{W} \sum_{c=1}^{C} (y - \hat{y})^2
\end{equation}

\begin{equation}
\text{MAE} = \frac{1}{N\cdot C \cdot H \cdot W} \sum_{i=1}^{N} \sum_{h=1}^{H} \sum_{w=1}^{W} \sum_{c=1}^{C} |y - \hat{y}|
\end{equation}

\begin{equation}
\text{SSIM(x,y)} = \frac{(2 \mu_x \mu_y + C_1)(2 \sigma_{xy} + C_2)}{(\mu_x^2 + \mu_y^2 + C_1)(\sigma_x^2 + \sigma_y^2 + C_2)}  
\end{equation}
\begin{equation}
\overline{\text{SSIM}} = \frac{1}{N} \sum_{i=1}^{N} \text{SSIM}(x,y)
\end{equation}

where:
\begin{align*}
N & \text{ is the number of images in the dataset,} \\
H & \text{ is the height of the images,} \\
W & \text{ is the width of the images,} \\
C & \text{ is the number of channels (e.g., 3 for RGB images),} \\
y & \text{ is the true pixel value at position } (i, h, w, c), \text{ and} \\
\hat{y} & \text{ is the predicted pixel value at position } (i, h, w, c).\\
\text{MAX} & \text{ is the maximum possible pixel value of the image (e.g., 255 for an 8-bit image),} \\
\text{MSE} & \text{ is the Mean Squared Error between the original and compressed image.}\\
\mu_x & \text{ is the average of } x, \\
\mu_y & \text{ is the average of } y, \\
\sigma_x^2 & \text{ is the variance of } x, \\
\sigma_y^2 & \text{ is the variance of } y, \\
\sigma_{xy} & \text{ is the covariance of } x \text{ and } y, \\
C_1 & = (K_1 L)^2 \text{ and } C_2 = (K_2 L)^2 \text{ are two variables to stabilize the division with weak denominator,} \\
L & \text{ is the dynamic range of the pixel values (typically, this is 255 for 8-bit images),} \\
K_1 & \text{ and } K_2 \text{ are small constants (typically, } K_1 = 0.01 \text{ and } K_2 = 0.03).\\
\end{align*}

\section{Additional Results} 

\paragraph{Probabilistic Forecasting For Vancouver Temperature:} \label{Van93results}
We now use the Vancouver93 dataset to build a forecasting model using SI. The approximating function we use is a residual-network based TCN inspired by \cite{bai2018empirical} along with time embedding on each layer. In this case we  propose a model that  uses a sequence of the last 5 days' temperatures to predict the next 5 days' temperatures.  For this problem, it is easily to visualize how SI learns the transportation from one distribution to another.

\begin{figure}[h]
    \centering
    \includegraphics[width=0.75\linewidth]{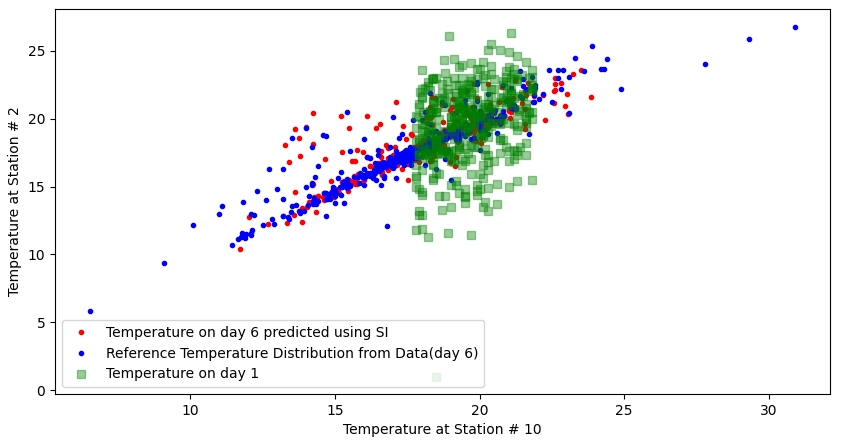}
    \caption{Phase diagram showing initial state and final states from data and model using SI after 5 days in between station 2 and station 10.}
    \label{fig:unconditionalVancouver93_5days}

    \centering
    \includegraphics[width=0.75\linewidth]{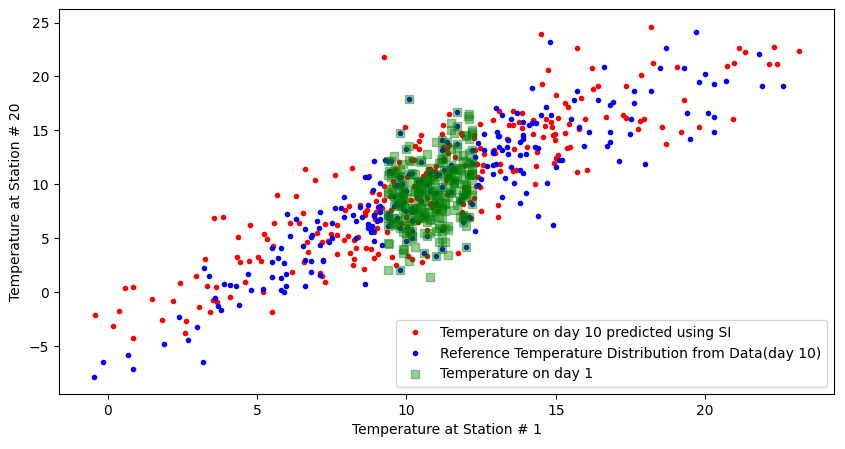}
    \caption{Phase diagram showing initial state and final states from data and model using SI after 10 days in between station 20 and station 1.}
    \label{fig:unconditionalVancouver93_10days}
\end{figure}

In the first experiment with this dataset, we take all the cases from the test-set where the temperature at Station 1 is 20$^\circ$C and we compare the actual distribution of temperatures at the same stations after 10 days. Essentially, if we look at the phase diagrams between any two stations for distributions of states after 10 days apart, we can visualize how the final distributions obtained using SI are very similar to the distributions we obtain from test data. Such distributions Often they look like skewed distributions, like we can see in figure \ref{fig:unconditionalVancouver93_5days} which shows the an extremely skewed distribution matching very well with the state after 5 days and similarly we can see in figure \ref{fig:unconditionalVancouver93_10days} which shows the distribution matching very well with the state after 10 days. Appendix \ref{app:van93stats} shows the matching histograms of the distributions for the later case. Some metrics for comparison of results can be seen in \cref{tab:AddstatsMetrics,tab:AddsimilarityMetricsMean,tab:AddsimilarityMetricsStd}.    

\begin{table}[!h]
\centering
\begin{tabular}{lcccc}
\toprule
 & {True}& {Our} & {True}& {Our}  \\
{Data}                 & {Mean}& {Mean} & {Std. Dev.}& {Std. Dev.}  \\\midrule
Vancouver93            & 1.85e1 & 1.89e1      & 4.05  & 3.97     \\
CloudCast              & 1.53e-2 & 1.51e-2     & 9.84e-2  & 9.84e-2          \\
\bottomrule
\end{tabular}
\caption{Comparison of mean and standard deviation of ensemble predictions.}
\label{tab:AddstatsMetrics}
\end{table}

\begin{table}[!h]
\centering
\begin{tabular}{lcccc}
\toprule
 {Data}                & {MSE}& {MAE} & {SSIM}\\\midrule
Vancouver93            & 5.9e-1 & 6.1e-1          & NA  \\
CloudCast              & 2.0e-7 & 6.0e-4          & 0.99  \\
\bottomrule
\end{tabular}
\caption{Similarity metrics for ensemble average.}
\label{tab:AddsimilarityMetricsMean}
\end{table}

\begin{table}[!h]
\centering
\begin{tabular}{lcccc}
\toprule
 {Data}                & {MSE}& {MAE} & {SSIM}\\\midrule
Vancouver93            & 1.8e-1 & 3.4e-1          & NA  \\
CloudCast              & 4.9e-7 & 6.0e-4          & 0.85  \\
\bottomrule
\end{tabular}
\caption{Similarity metrics for ensemble standard deviation.}
\label{tab:AddsimilarityMetricsStd}
\end{table}

\section{Statistical Comparison of Our Ensemble Predictions} 
\subsection{Predator-Prey Model} \label{app:predpreystats}
\begin{figure*}[!h]
        \centering
        \begin{tabular}{cc}
        \includegraphics[width=0.5\linewidth]{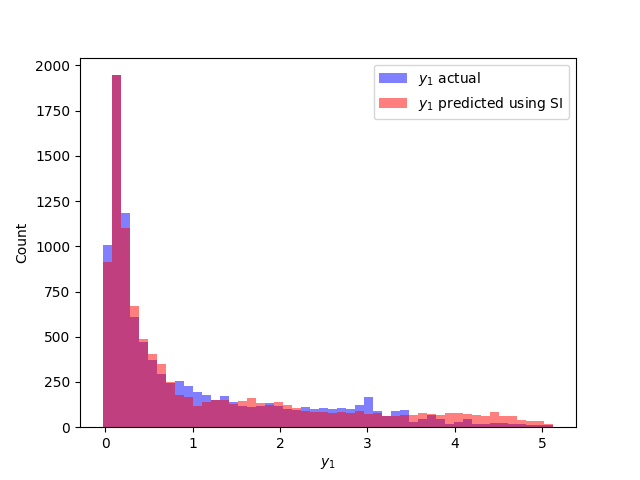} &
        \includegraphics[width=0.5\linewidth]{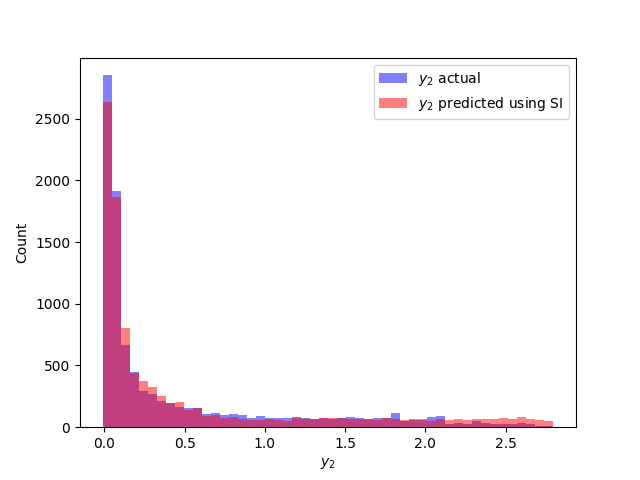}
        \end{tabular}
        \caption{Histograms of actual final distribution of $\bfy_1$ (left) and $\bfy_2$ (right) compared with that obtained using SI on the predator-prey model}
        \label{fig:hist_predprey}
\end{figure*}

Figures in \ref{fig:hist_predprey} shows the histograms of $\bfy_1$ and $\bfy_2$ respectively. It can be noticed that the histogram of an actual state variable is matching very well with the histogram obtained from our results using SI.

\subsection{Vacouver93} \label{app:van93stats}
\begin{figure}[!h]
    \centering
    \includegraphics[width=1.0\linewidth]{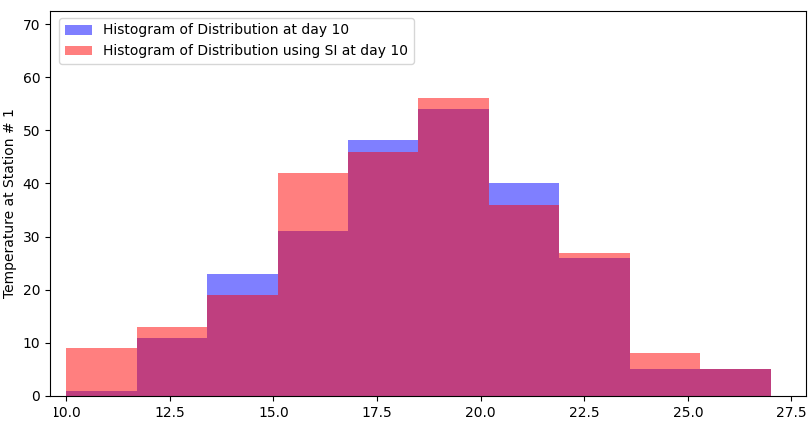}
    \caption{Histograms of distributions observed after 10 days and distribution obtained using SI excluding outliers.}
    \label{fig:unconditionalVancouver93station1Hist}
\end{figure}

Figure \ref{fig:unconditionalVancouver93station1Hist} shows the histograms of the two distributions which are very similar to suggest that SI efficiently learns the transport map to the distribution of temperatures at a station for this problem whose deterministic solution is very tough. 

\begin{figure}[!h]
    \centering
    \includegraphics[width=1.0\linewidth]{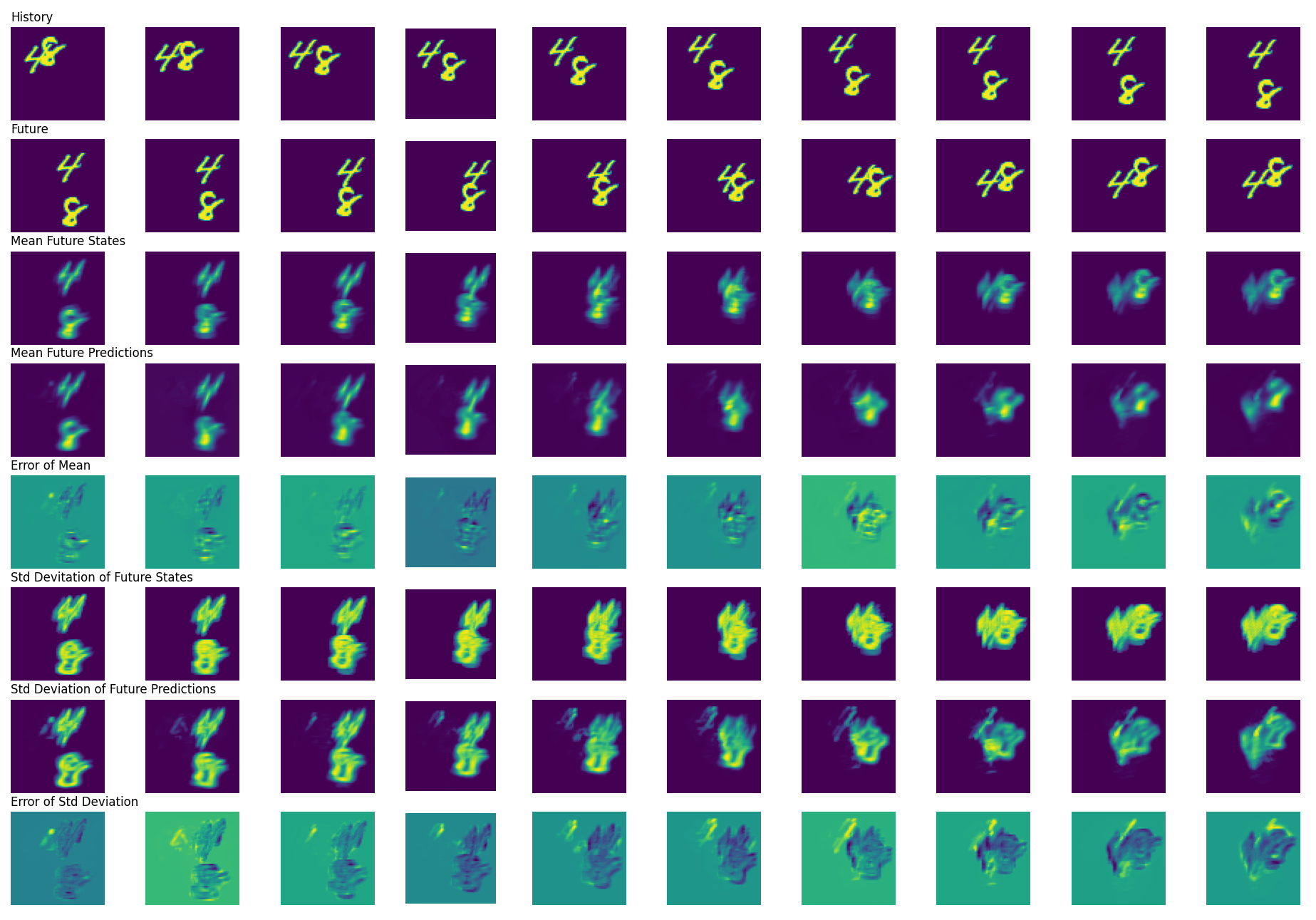}
    \caption{Statistical comparison of Moving MNIST sequences predicted using SI with random perturbed initial states.}
    \label{fig:MMLowPert}
\end{figure}

\begin{figure}[!h]
    \centering
    \includegraphics[width=1.0\linewidth]{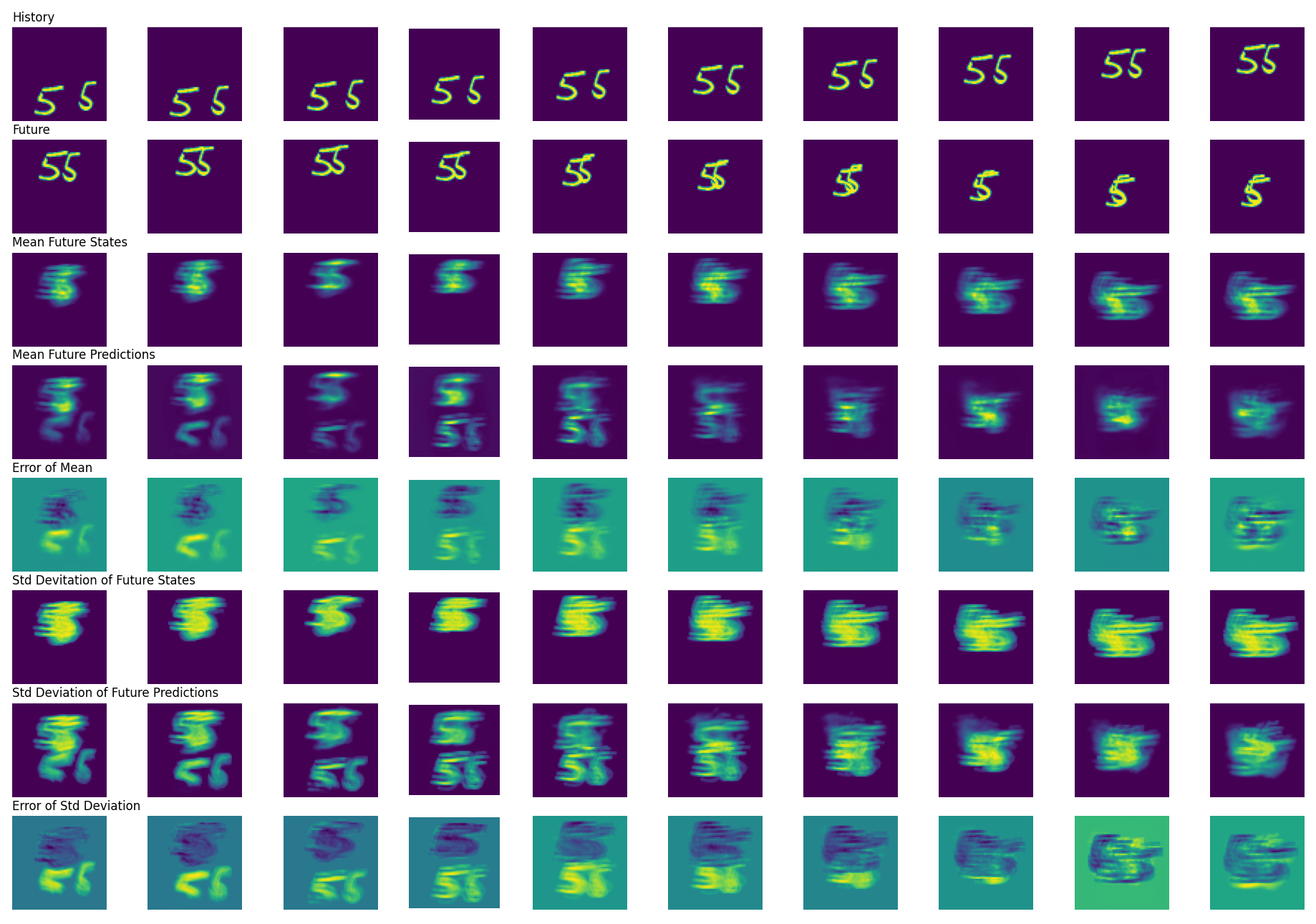}
    \caption{Statistical comparison of Moving MNIST sequence with random highly perturbed initial states.}
    \label{fig:MMLargePert}
\end{figure}

\subsection{Moving MNIST} \label{app:MMstats}
 Let us take as input, initial 10 frames from a case in Moving MNIST test set as history and actual 10 subsequent frames in the sequence as future. For such a case Figures \ref{fig:MMLowPert} and \ref{fig:MMLargePert} show statistical images for small perturbation and large perturbation respectively. 100 perturbed samples are taken and their final states are predicted using SI. The figures showcase how similar the ensemble mean and the ensemble standard deviations are. This is a fair justification that the uncertainty designed on Moving MNIST is captured very well. 

\subsection{Cloudcast}\label{app:CCstats}
\subsubsection{Predictions}
\begin{figure}[!h]
    \centering
    \includegraphics[width=1.0\linewidth]{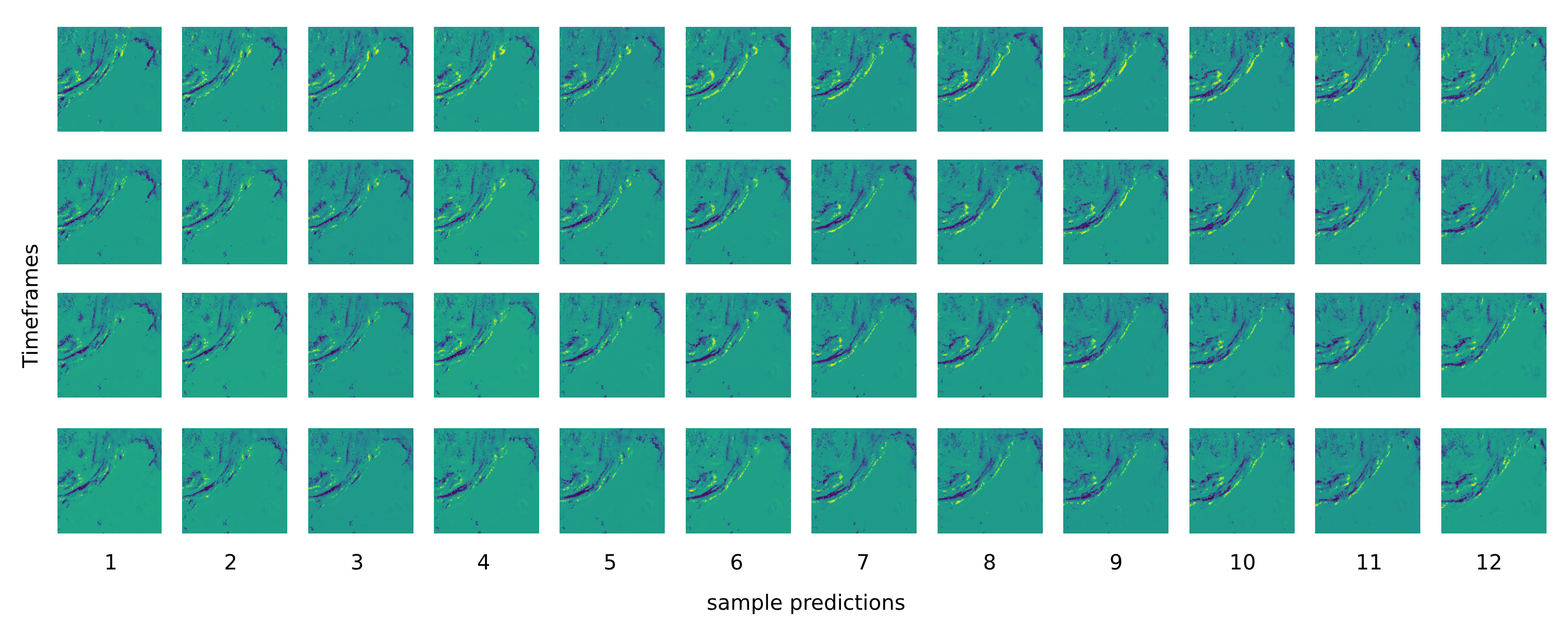}
    \caption{Random predictions of differential change from a single sequence from cloudcast testset.}
    \label{fig:cloudcastensembleupdatepredictions}
\end{figure}

\begin{figure}[!h]
    \centering
    \includegraphics[width=1.0\linewidth]{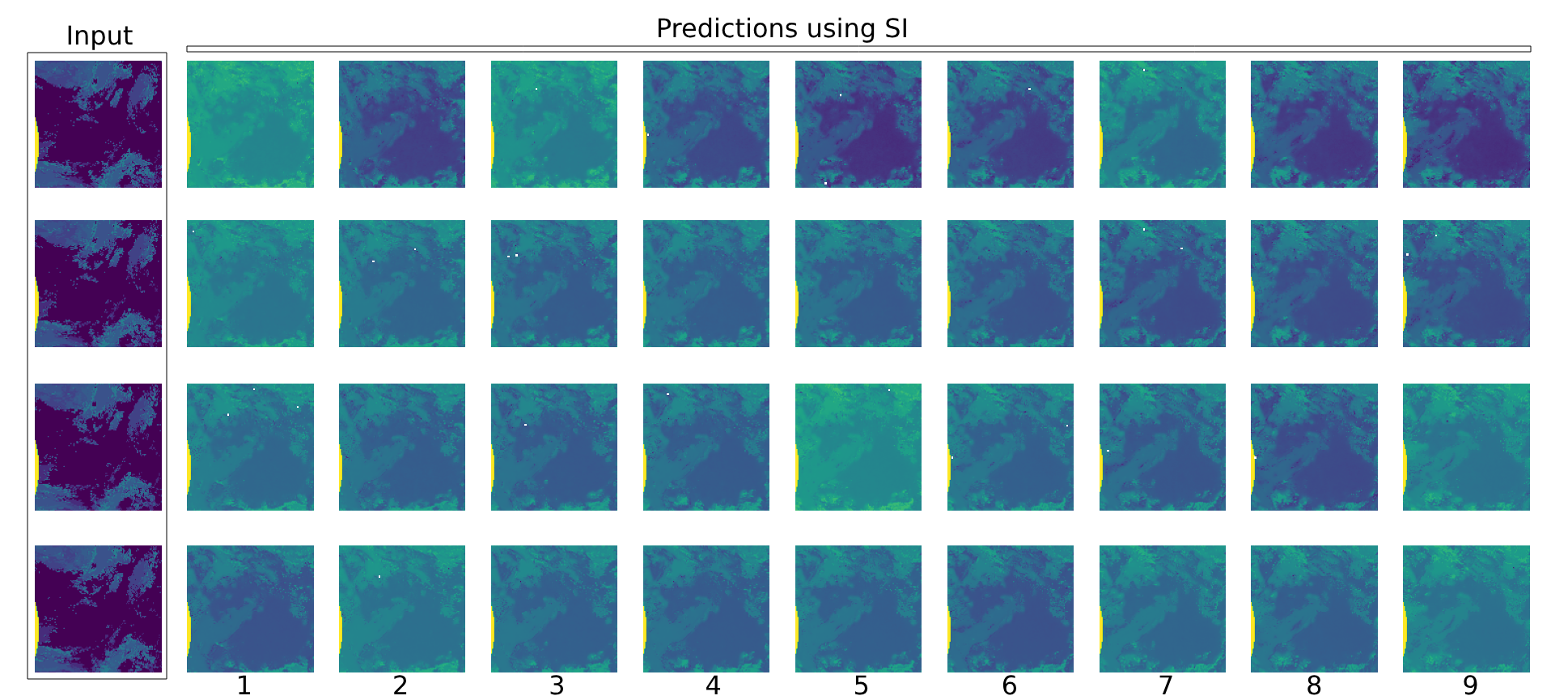}
    \caption{Random predictions from a single sequence from CloudCast testset for prediction after 3 hours.}
    \label{fig:cloudcastensemblepredictions}
\end{figure}

As per the default configuration of 4 timeframe sequences from the test set of Cloudcast dataset is used to predict 4 subsequent timeframes in the future. A point to notice is each time step equals to 15 minutes in real time and hence the differential changes is what we observe. For one such a case Figure \ref{fig:cloudcastensembleupdatepredictions} shows twelve different predictions. They are very similar but not the same once we zoom into high resolution. Figure \ref{fig:cloudcastensemblepredictions} shows the prediction after 3 hours by autoregression to showcase how the clouds look noticeably different. The small cloud patches are visibly different in shape and sizes. Cloud being a complex and nonlinear dynamical system, the slight non-noticeable difference in 1 hour can lead to very noticeably different predictions after 3 hours.   

\subsubsection{Statistical comparison}
\begin{figure}[!h]
    \centering
    \includegraphics[width=0.85\linewidth]{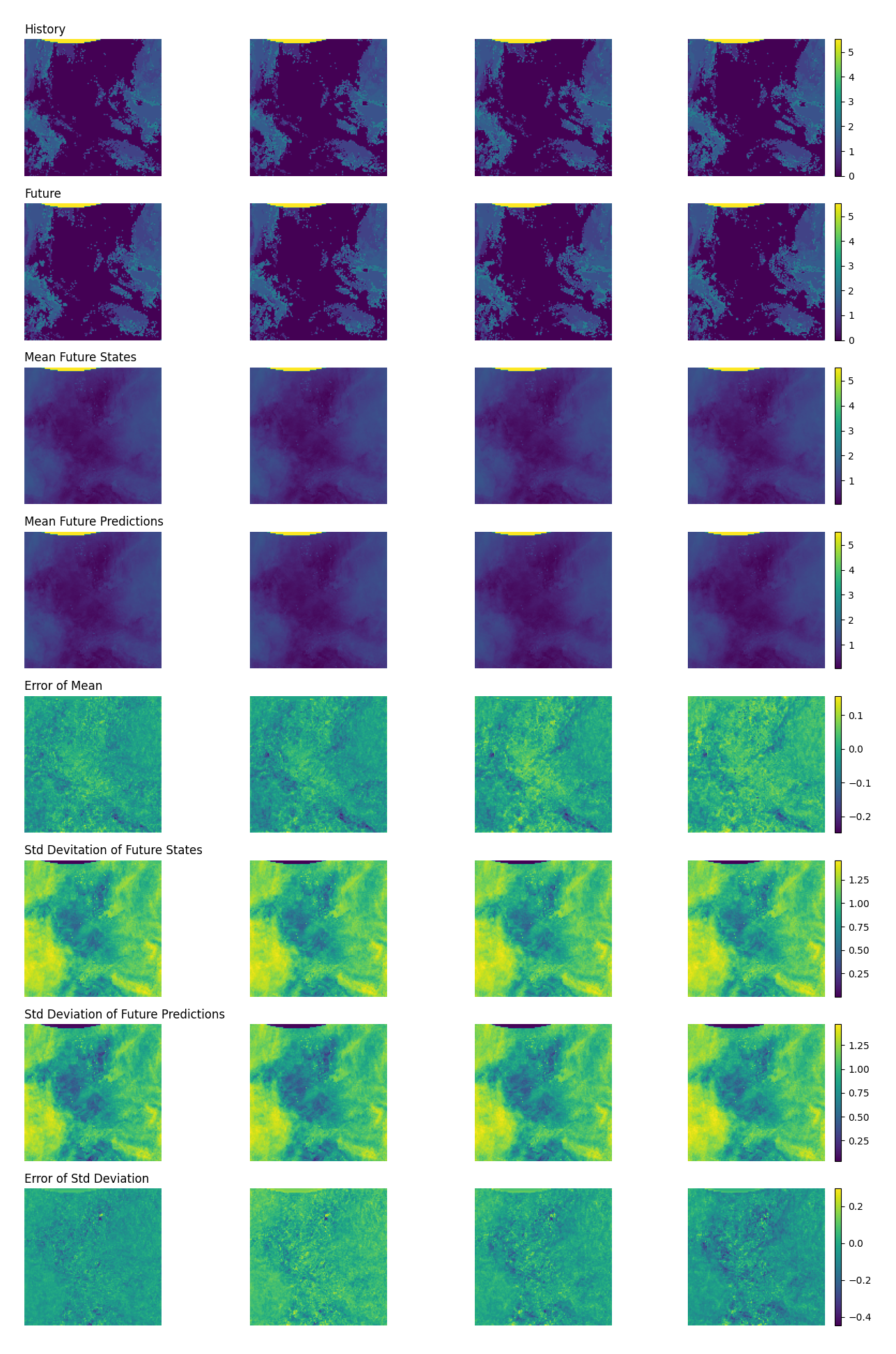}
    \caption{Statistical comparison of CloudCast sequences predicted using SI for a single initial state.}
    \label{fig:CCPertStats}
\end{figure}

Figure \ref{fig:CCPertStats} shows statistical images for generated outputs on CloudCast testset using SI. Similar samples are taken as the set of initial states to predict their final states using SI. The figures showcase how similar the ensemble mean and the ensemble standard deviations are. 

\subsection{WeatherBench}\label{app:wBstats}
\subsubsection{Predictions}
\begin{figure*}[!h]
    \centering
    \includegraphics[width=1.0\linewidth]{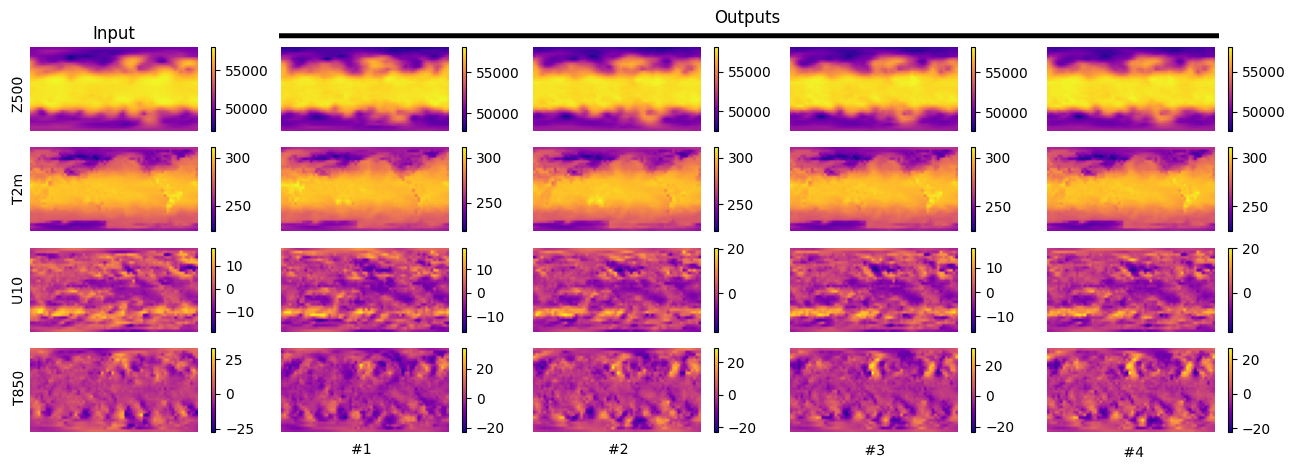}
    \caption{Four sample stochastic forecasts of Z500, T2m, U10 and T850 after 2 days obtained using SI.}
    \label{fig:wBsample20forecasts}
\end{figure*}

Figure \ref{fig:wBsample20forecasts} shows some of predictions using SI with mild perturbations.
\subsubsection{Statistical comparison}
\begin{figure*}[!h]
    \centering
    \includegraphics[width=1.0\linewidth]{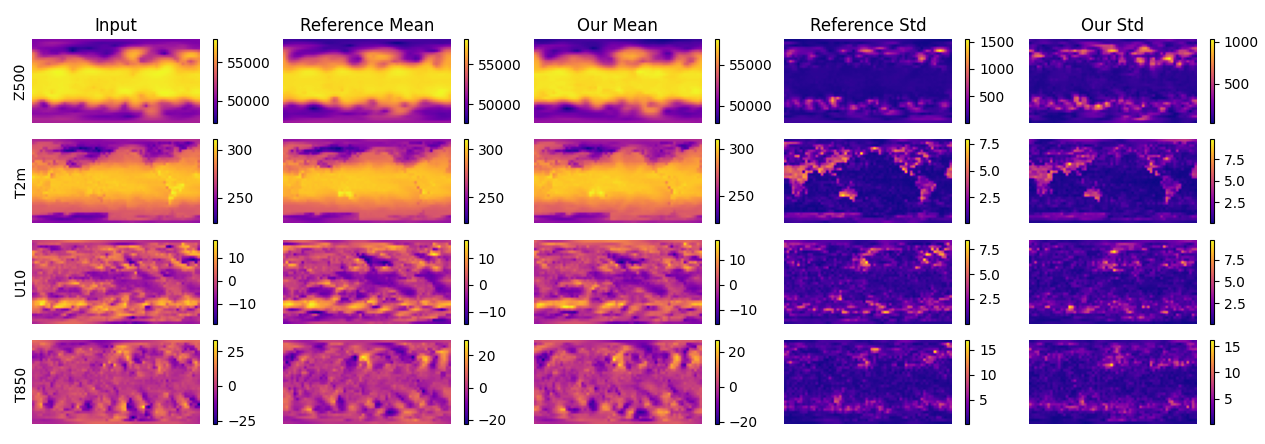}
    \caption{Statistical comparisons of Z500, T2m, U10 and T850 for 2 day ensemble forecasting using SI using 20 mildly perturbed samples.}
    \label{fig:wBstats20forecasts}
\end{figure*}

\begin{figure*}[!h]
    \centering
    \includegraphics[width=1.0\linewidth]{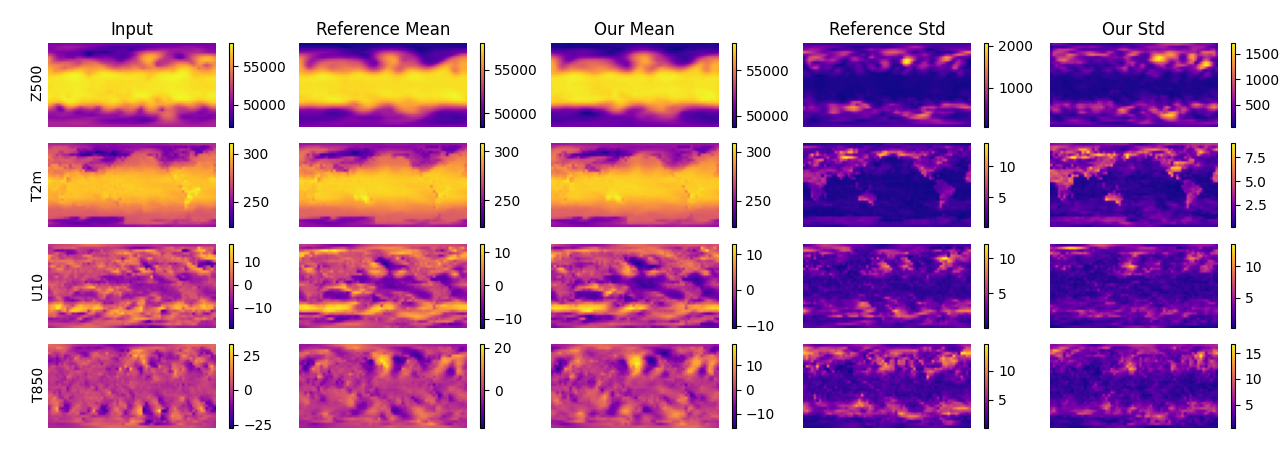}
    \caption{Statistical comparisons of Z500, T2m, U10 and T850 for 2 day ensemble forecasting using SI using 78 strongly perturbed samples.}
    \label{fig:wBstats78forecasts}
\end{figure*}

Figure \ref{fig:wBstats20forecasts} shows the statistical comparison of ensemble predictions with 20 mildly perturbed states. Figure \ref{fig:wBstats78forecasts} shows the statistical comparison of ensemble predictions with 78 strongly perturbed states. The mean and standard deviation of the states for 2 day (48 hours predictions) can be easily compared. Table \ref{tab:similarityMetricsWB6hrs} and \cref{tab:similarityMetricsWB2days} shows the metrics to compare the accuracy of our ensemble mean and ensemble standard deviation for 6 hour and 2 day predictions respectively. 


\begin{table*}[!h]
\centering
\begin{tabular}{l|ccccc|ccccc}
\toprule
& \multicolumn{5}{c}{Ensemble Mean} & \multicolumn{5}{|c}{Ensemble Std. Dev.} \\ \cline{2-11}
{Variable} &   True Score    &   Our Score    &   MSE($\downarrow$)   &   MAE($\downarrow$)   &   SSIM($\uparrow$)   &   True Score  &  Our Score     &   MSE($\downarrow$)    &   MAE($\downarrow$)    &   SSIM($\uparrow$)    \\ \hline
Z500       &  5.40e4   &  5.40e4  &  2.00e4 &  1.04e2    &  0.992    & 3.96e2 & 3.74e2 &  1.83e4
&  9.16e1    &  0.850 \\
T2m        &   2.78e2  &  2.78e2  &  2.62   &  9.33e-1    &   0.986  & 1.78 & 1.84   &       8.90e-1 &  5.60e-1   &  0.745\\
U10        &  -1.85e-1 & -2.70e-1 &  1.27   &  8.40e-1   &    0.886  & 2.51  & 2.34  &        1.12   & 7.35e-1 &  0.708 \\
T850       &  -2.43e-2  & -3.74e-2 &  1.62  &  9.79e-1  &    0.812  & 3.75   & 3.58   &       1.29  &  8.26e-1    & 0.774\\
\bottomrule
\end{tabular}
\caption{Similarity metrics for weatherBench ensemble prediction after 6 hours.}
\label{tab:similarityMetricsWB6hrs}
\end{table*}

\begin{table*}[!h]
\centering
\begin{tabular}{l|ccccc|ccccc}
\toprule
& \multicolumn{5}{c}{Ensemble Mean} & \multicolumn{5}{|c}{Ensemble Std. Dev.} \\ \cline{2-11}
{Variable} &   True Score    &   Our Score    &   MSE($\downarrow$)   &   MAE($\downarrow$)   &   SSIM($\uparrow$)   &   True Score  &  Our Score     &   MSE($\downarrow$)    &   MAE($\downarrow$)    &   SSIM($\uparrow$)    \\ \hline
Z500       &  5.40e4   &  5.40e4  &  7.86e4 &  2.04e2    &  0.978    & 3.58e2 & 3.78e2 &       4.52e4 &  1.41e2    &  0.630\\
T2m        &   2.78e2  &  2.78e2  &  3.05   &  1.06    &    0.986  & 1.83   & 1.95   &       1.54   &  7.3e-1    &  0.681\\
U10        &  -1.16e-1 & -1.10e-1 &  2.42   &  1.17    &    0.820  & 2.42   & 2.58   &        2.19   & 1.04    &  0.561\\
T850       & -8.09e-2  & -8.12e-2 &  4.30   &  1.53    &    0.688 & 3.64   & 3.72   &       3.12   &  1.28    & 0.561\\
\bottomrule
\end{tabular}
\caption{Similarity metrics for weatherBench ensemble prediction after 2 days.}
\label{tab:similarityMetricsWB2days}
\end{table*}

\section{Perturbations using SI: Additional Study} \label{app:MM} 
\subsection{MovingMNIST} 
\begin{figure}[h]
    \centering
    \includegraphics[width=1.0\linewidth]{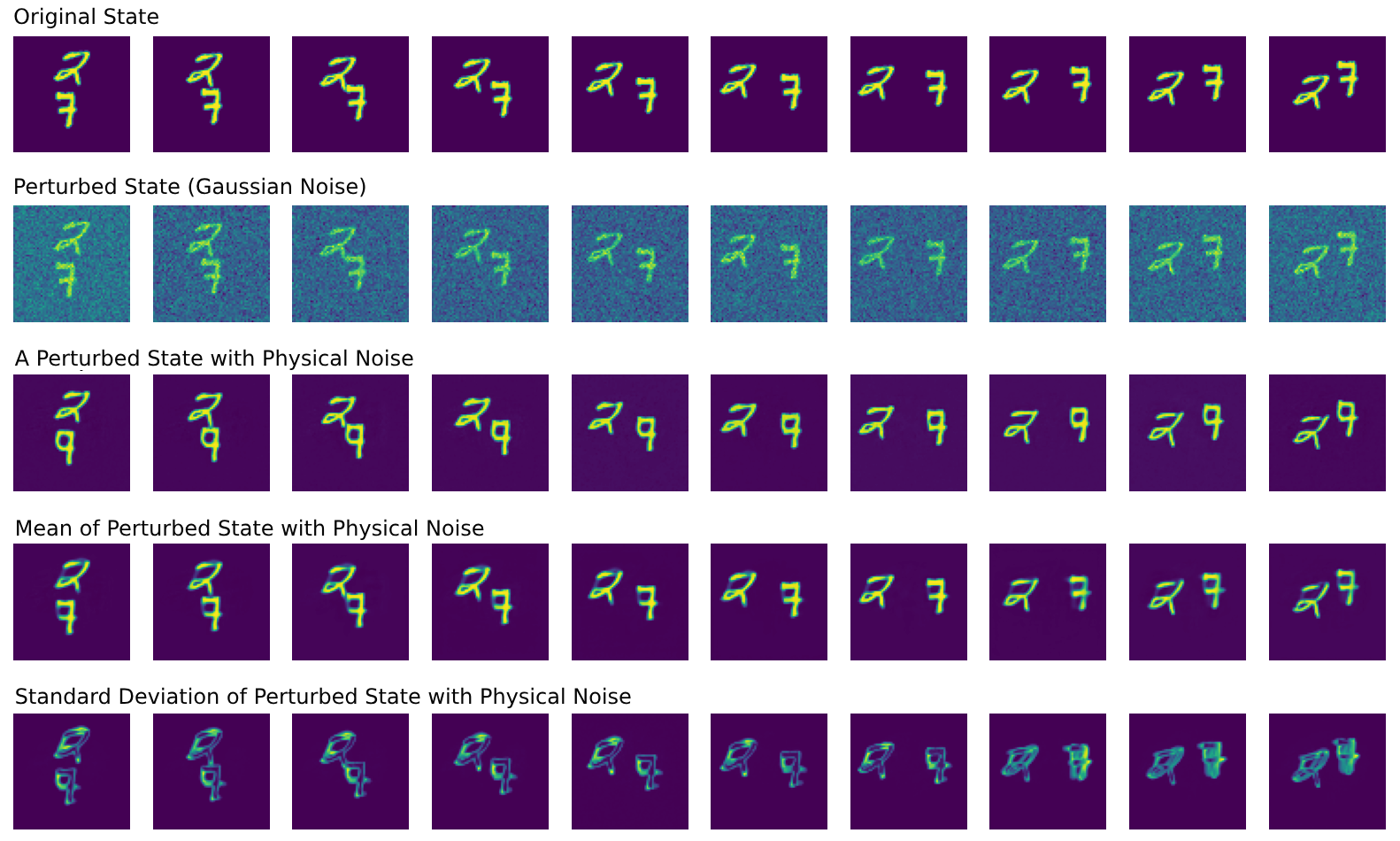}
    \caption{Statistical comparison of 100 perturbed states from a single MovingMNIST sequence using SI.}
    \label{fig:MMPertStats}
\end{figure}

\begin{figure}[h]
    \centering
    \includegraphics[width=1.0\linewidth]{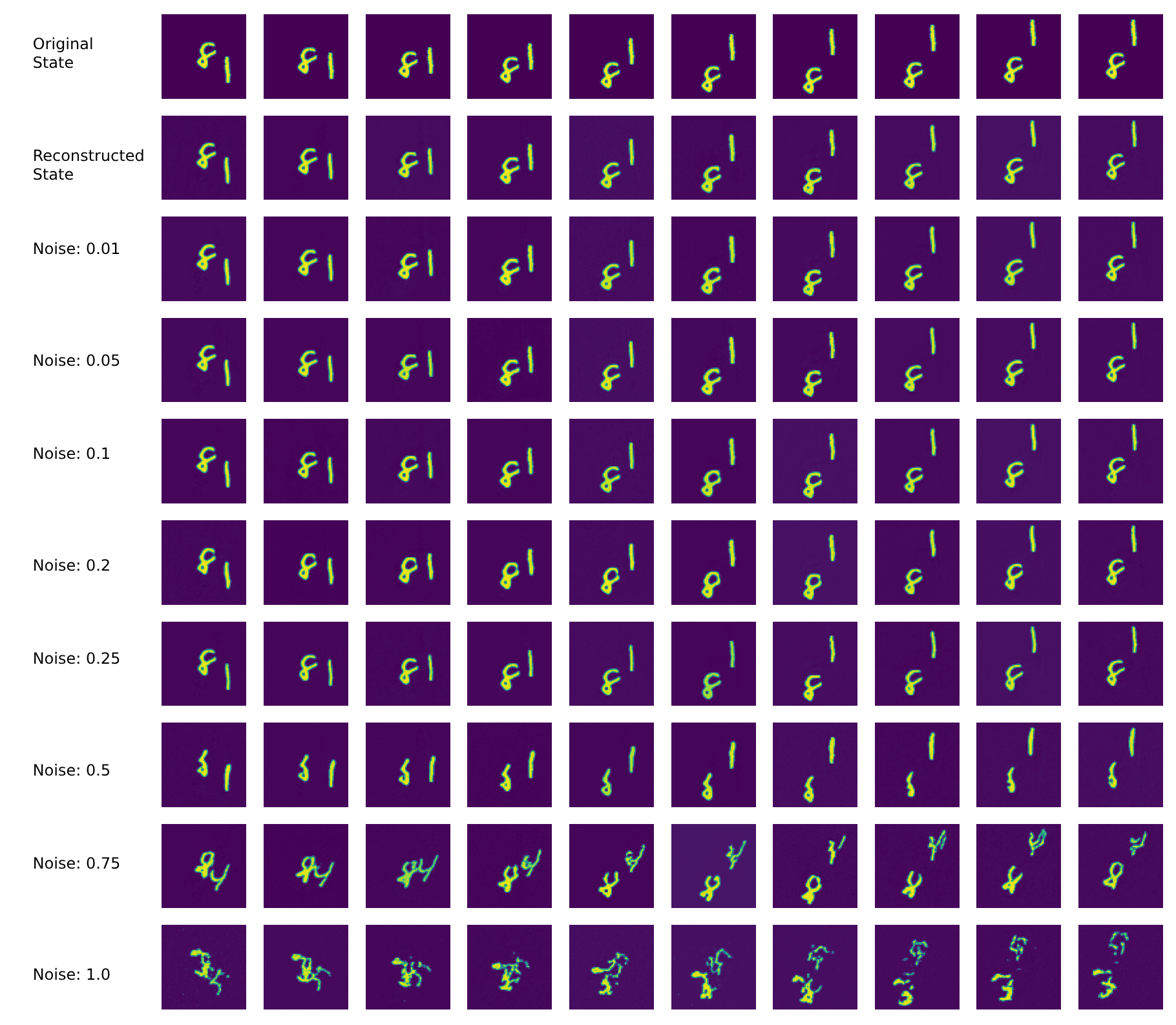}
    \caption{Perturbed states for a single MovingMNIST sequence using SI for different levels of noise.}
    \label{fig:MMPertvsNoise}
\end{figure}

Figure in \ref{fig:MMPertStats} shows how the perturbed state generated using SI is better than Gaussian perturbations. Also, a crucial factor for the sensitivity of perturbation is showcased, where a generated sequence from MovingMNIST has a different digit is shown in the third row of the figure. With mild perturbaton, however, the mean of the 100 samples matches very well with the original state. The standard deviation of those states show the scope of perturbation, which is good. Figure \ref{fig:MMPertvsNoise} shows how a perturbed state varies with different noise levels. With larger noise, the digits seems to transform, like 'one' turns to 'four', 'eight' turns to 'three', and so on. The model understands than on transitioning, the digits should turn into another digit.  

\subsection{WeatherBench} 
\begin{figure}[h]
    \centering
    \includegraphics[width=1.0\linewidth]{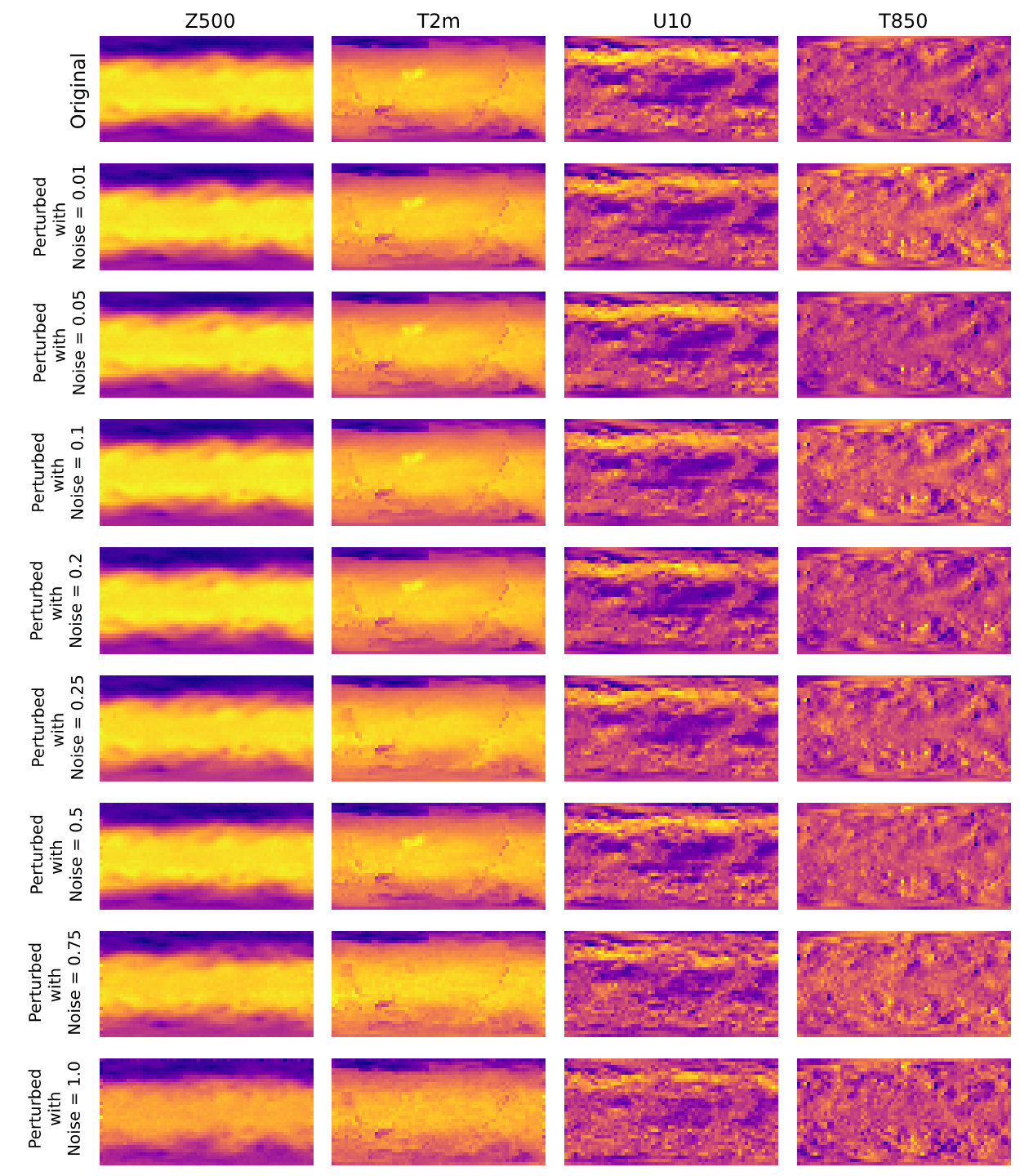}
    \caption{Perturbed states for a WeatherBench state using SI for different levels of noise.}
    \label{fig:wBPertvsNoise}
\end{figure}

Figure in \ref{fig:wBPertvsNoise} shows the four curated important variables for weather prediction, Z500, T2m, U10 and T850, perturbed with different values of noise.

% % NOTE: necessary when ptmx or no mathfont class option is given
% \providecommand{\upGamma}{\Gamma}
% \providecommand{\uppi}{\pi}

\section{Hyperparameter Settings and Computational Resources}
\subsection{UNet Training}
Table \ref{tab:hyperparameters_mnistCC} shows the hyperparameter settings for training on MovingMNIST and CloudCast datasets. While table \ref{tab:hyperparameters_wB} shows the settings for training on WeatherBench.

\label{UNetsettings}
\begin{table}[!h]
\centering
\begin{tabular}{lcccc}
\toprule
{Hyperparameter}     & {Symbol} & {Value}  \\ \midrule
Learning Rate               & $\eta$          & $1e-04$         \\
Batch Size                  & $B$             & $64$           \\
Number of Epochs            & $N$             & $200$           \\
Optimizer                   & -               & Adam            \\
Dropout                     & -               & 0.1             \\
Number of Attention Heads   & -               & 4             \\
Number of Residual Blocks   & -               & 2             \\
\bottomrule
\end{tabular}
\caption{Neural Network Hyperparameters for training Moving MNIST and CloudCast}
\label{tab:hyperparameters_mnistCC}
\end{table}

\label{UNetsettingswB}
\begin{table}[!h]
\centering
\begin{tabular}{lcccc}
\toprule
{Hyperparameter}     & {Symbol} & {Value}  \\ \midrule
Learning Rate               & $\eta$          & $1e-04$         \\
Batch Size                  & $B$             & $8$           \\
Number of Epochs            & $N$             & $50$           \\
Optimizer                   & -               & Adam            \\
Dropout                     & -               & 0.1             \\
Number of Attention Heads   & -               & 4             \\
Number of Residual Blocks   & -               & 2             \\
\bottomrule
\end{tabular}
\caption{Neural Network Hyperparameters for training WeatherBench}
\label{tab:hyperparameters_wB}
\end{table}

\subsection{Computational Resources} 
\label{resources}
All our experiments are conducted using an NVIDIA RTX-A6000 GPU with 48GB of memory. 

\end{document}